\documentclass[12pt, notitlepage,aps,preprintnumbers,superscriptaddress,nofootinbib,aps,prd,floatfix]{revtex4-2}
\pdfoutput=1
\usepackage{float}
\usepackage{physics,slashed}
\usepackage{amsfonts,amsmath,amssymb}
\usepackage{mathrsfs}
\usepackage{bm}
\usepackage{epsfig}
\usepackage{graphicx}   
\usepackage{subfigure}  
\usepackage{verbatim}   
\usepackage{color}      
\usepackage{hyperref}   
\usepackage{url}
\usepackage{xcolor}
\definecolor{nicered}{rgb}{0.5,0.,0.}
\definecolor{nicegreen}{rgb}{0.,0.5,0.}
\definecolor{niceblue}{rgb}{0.,0.,0.5}
\hypersetup{colorlinks,citecolor=nicegreen,linkcolor=nicered,urlcolor=niceblue}
\usepackage{orcidlink}

\usepackage{array}
\usepackage{dcolumn}
\usepackage{multirow}
\usepackage{cprotect}
\usepackage{framed}
\usepackage{setspace}
\usepackage{enumitem}
\setlist{nolistsep} 
\usepackage{adjustbox}
\usepackage{feynmf}
\usepackage{comment}

\usepackage{tikzsymbols}
\newcommand{\GeV}{\textrm{GeV}}
\newcommand{\TeV}{\textrm{TeV}}
\newcommand{\calB}{\mathcal{B}}
\newcommand{\calL}{\mathcal{L}}

\newcommand{\calM}{\mathcal{M}}
\newcommand{\calT}{\mathcal{T}}
\newcommand{\calS}{\mathcal{S}}
\newcommand{\hc}{\textrm{h.c.}}

\newcommand{\ie}{\emph{i.e.}}
\newcommand{\br}{\mathcal{B}}
\newcommand{\met}{\ensuremath{\slashed{E}_T}}
\newcommand{\pslash}{\not\! p}

\def\lsim{\raise0.3ex\hbox{$\;<$\kern-0.75em\raise-1.1ex\hbox{$\sim\;$}}}
\def\gsim{\raise0.3ex\hbox{$\;>$\kern-0.75em\raise-1.1ex\hbox{$\sim\;$}}}

\begin{document}
\title{Collider probes of baryogenesis with maximal CP asymmetry}
\author{Debasish Borah}
\email{dborah@iitg.ac.in}
\affiliation{Department of Physics, Indian Institute of Technology Guwahati, Assam 781039, India}
\author{Kun Cheng}
\email{kun.cheng@pitt.edu}
\affiliation{Pittsburgh Particle Physics, Astrophysics, and Cosmology Center, Department of Physics and Astronomy, University of Pittsburgh, Pittsburgh, PA 15260, USA\looseness=-1}
\author{Arnab Dasgupta}
\email{arnabdasgupta@pitt.edu}
\affiliation{Pittsburgh Particle Physics, Astrophysics, and Cosmology Center, Department of Physics and Astronomy, University of Pittsburgh, Pittsburgh, PA 15260, USA\looseness=-1}
\author{Tao Han}
\email{than@pitt.edu}
\affiliation{Pittsburgh Particle Physics, Astrophysics, and Cosmology Center, Department of Physics and Astronomy, University of Pittsburgh, Pittsburgh, PA 15260, USA\looseness=-1}
\author{Keping Xie}
\email{xiekepi1@msu.edu}
\affiliation{Pittsburgh Particle Physics, Astrophysics, and Cosmology Center, Department of Physics and Astronomy, University of Pittsburgh, Pittsburgh, PA 15260, USA\looseness=-1}
\affiliation{Department of Physics and Astronomy, Michigan State University, East Lansing, MI 48824, USA\looseness=-1}
	
\begin{abstract}
We propose a novel collider probe of baryogenesis at TeV scale by measuring decay asymmetries into particle and anti-particle final states. Motivated by the idea of Dirac leptogenesis, we consider an extension of the standard model with new colored and $SU(2)_L$ singlet particles in such a way that the out-of-equilibrium decay of heavy colored fermions creates equal and opposite CP asymmetries in two sectors, prevented from equilibrating with each other. While the TeV scale viability of this mechanism requires a resonantly enhanced CP asymmetry, the latter also plays a crucial role leading to observable decay asymmetries in colliders. In addition to discussing conventional signatures of such heavy colored particles, namely, mono-jet plus missing transverse energy, displaced vertex, colored track at hadron colliders, we also show the unique possibility of measuring decay asymmetries via forward-backward and charge asymmetries at future muon colliders. In addition to being a verifiable TeV-scale baryogenesis scenario, the model also predicts a singlet scalar dark matter candidate consistent with the required thermal dark matter properties near the Higgs resonance.
\end{abstract}

\preprint{MSUHEP-25-004, PITT-PACC-2504}
  
\maketitle
\tableofcontents

\section{Introduction}
The observed baryon and anti-baryon asymmetry of the present Universe (BAU) has been one of the most eminent and longstanding puzzles in particle physics and cosmology \cite{ParticleDataGroup:2020ssz, Planck:2018vyg}. This observed excess of baryons over anti-baryons is quantified in terms of the baryon to photon ratio as \cite{Planck:2018vyg} 
\begin{equation}
\eta_B = \frac{n_{B}-n_{\bar{B}}}{n_{\gamma}} \simeq (6.12\pm 0.04) \times 10^{-10}, 
\label{etaBobs}
\end{equation} 
based on cosmic microwave background (CMB) measurements, which also agrees well with the big bang nucleosynthesis (BBN) estimates \cite{ParticleDataGroup:2020ssz}. The necessary conditions for dynamically generating a nonzero baryon asymmetry, laid out by Sakharov \cite{Sakharov:1967dj}, cannot be satisfied in the required amount in the standard model (SM) of particle physics, leading to several proposals beyond the standard model (BSM). Among these, baryogenesis~\cite{Ellis:1978xg, Yoshimura:1978ex, Weinberg:1979bt, Kolb:1979qa, Fry:1980bd} and leptogenesis~\cite{Fukugita:1986hr} have been the most studied. Although typical baryogenesis scenarios correspond to a very high scale, out of reach from direct searches in laboratory experiments, low- or intermediate-scale baryogenesis scenarios \cite{Babu:2006xc,Babu:2006wz,Allahverdi:2010im, Allahverdi:2010rh, Allahverdi:2013tca, Allahverdi:2013mza,Allahverdi:2017edd, Arcadi:2015ffa} involving new colored particles can be probed via neutron-antineutron oscillations $(n-\bar{n})$ \cite{Baldo-Ceolin:1994hzw, Super-Kamiokande:2011idx, SNO:2017pha, Fridell:2021gag} or at collider experiments such as the large hadron collider (LHC) \cite{Baldes:2011mh}. These scenarios typically consider $\Delta (B-L)=2$ violations which evade strong proton decay bounds on $\Delta (B-L)=1$ processes. In addition, a nonzero $B-L$ number can not be washed out by electroweak sphalerons \cite{Kuzmin:1985mm} that conserve $B-L$ and violate $B+L$. Similarly, the scale of leptogenesis can also vary from high to low scale, although most of these scenarios are very difficult to probe directly \cite{Chun:2017spz}. Baryogenesis models involving new colored particles are more promising for direct probe if such particles are kinematically accessible at collider energies. However, constraints on baryon number violation (BNV) from nucleon decay and $n-\bar{n}$ oscillation limits push such scenarios to a tight corner in the theory parameter space unless some specific flavor structure is chosen.

Motivated by the idea of Dirac leptogenesis \cite{Dick:1999je,Murayama:2002je} which relies on equal and opposite CP asymmetries in left- and right-chiral lepton sectors without any net lepton number violation, we consider a similar setup where a net baryon asymmetry in the visible sector can survive. While such a measure is optimized by considering resonantly enhanced CP asymmetries, the stringent limits on low-scale BNV are circumvented by considering a Dirac baryogenesis setup without any net BNV. Due to the absence of net BNV, equal and opposite CP asymmetries prevail in two sectors that are prevented from equilibrating. To achieve this goal, the SM field content is extended by two copies of a vector-like colored fermion $\psi_{i}\ (i=1,2)$, a chiral gauge singlet fermion $N$, one colored scalar $\eta$, and one singlet scalar $\phi$. A discrete unbroken $Z_4$ symmetry, acting non-trivially on $\psi_i, \eta$ and $\phi$, is imposed to keep the unwanted terms away. The out-of-equilibrium decay of the vector-like colored fermion sources the baryon asymmetry. While CP asymmetries in the two sectors are equal and opposite by unitarity criteria, we utilize phase-space limits to suppress the baryon asymmetry stored in the new colored scalar $\eta$ to minimize the washout of the SM baryon asymmetry due to the late decay of $\eta$. Due to the strong dilution of the asymmetry induced by strong interactions of $\psi_{i}$, resonantly enhanced CP asymmetry is considered. Such near-maximal CP asymmetry is necessary to achieve TeV scale baryogenesis and will lead to a significant difference in the decay of the new colored particle and its anti-particle, potentially yielding a unique and spectacular experimental signature. Although the complex singlet scalar $\phi$ has been included primarily to realize Dirac baryogenesis, it naturally emerges as a cold dark matter (DM) candidate, stabilized by the symmetry $Z_4$.

With the above-mentioned particle physics setup as a working example, we propose a novel probe of low-scale baryogenesis at colliders by measuring the asymmetry in decays of heavy colored particles into particle and anti-particle, a basic ingredient of any baryogenesis scenario. 
We show the viability of the model in generating the correct baryon asymmetry of the Universe at TeV scale and explore the detection prospects at present and future colliders, including the LHC and future high-energy lepton colliders, via the di-jet plus missing energy signatures, displaced vertex, and as well as forward-backward and charge asymmetry.

This paper is organized as follows. In section \ref{sec:model}, we describe our model of Dirac baryogenesis followed by the discussion of baryogenesis in section \ref{sec:cosmo}. In section \ref{sec:collider}, we discuss the details of collider signatures and finally conclude in section \ref{sec:conclusion}.
We defer the details of the loop calculations in App.~\ref{app:Asym} and the interference effects in App.~\ref{app:BreitWigner}.

\section{The Model}
\label{sec:model}
The SM is extended by a set of new particles shown in Table \ref{tab:particle} along with their quantum numbers under the symmetry of the model. An additional $Z_4$ symmetry ($i^4=1$) is considered in order to avoid unwanted terms in the Lagrangian. The SM fields transform trivially under this $Z_4$ symmetry. The relevant part of the Lagrangian involving BSM fields can be written as
\begin{equation}\label{eq:Lag}
  -\mathcal{L}_{\rm BSM} \supset y_L \bar{\psi}_L u_R \phi + y_R \bar{\psi}_R N \eta + M_{\psi} \bar{\psi}_L\psi_R + \frac{1}{2}M_N \bar{N}^c N + {\rm h.c.}
\end{equation}
One can assign baryon number $1/3$ each to $\psi, \eta$ such that the above Lagrangian does not violate baryon number. However, in the spirit of Dirac baryogenesis, we can still generate equal and opposite CP asymmetries in two sectors, namely, $u_R \phi$ and $N \eta$. Due to net baryon number conservation, the CP asymmetries resulting from $\psi \rightarrow u \phi$ and $\psi \rightarrow N \eta$ turn out to be equal and opposite. As long as $u_R \phi \leftrightarrow N \eta$ does not enter equilibrium, the asymmetry in the SM quark sector can survive, leading to the observed baryon asymmetry. As we discuss below, using phase space kinematics, it is also possible to generate different baryon asymmetries in $u_R \phi$ and $N \eta$ final states even though the corresponding CP asymmetries are exactly equal and opposite. This leads to the survival of a net asymmetry even after the late decay of $\eta$ into SM quarks. 

Since $Z_4$ symmetry remains unbroken, the model also predicts a stable particle. We choose this to be the singlet scalar $\phi$, the lightest particle with non-trivial charge under $Z_4$. This complex scalar singlet can be a dark matter candidate whose relic density will be governed by Higgs portal and Yukawa portal interactions. The singlet chiral fermion $N$, uncharged under $Z_4$ can couple to the SM lepton doublet, Higgs and generate light neutrino masses via type-I seesaw mechanism \cite{Minkowski:1977sc,Yanagida:1979as,Gell-Mann:1979vob,Mohapatra:1980yp, Schechter:1980gr}. Given that the SM can not explain the observed neutrino mass and mixing, the model also provides a way to solve this puzzle.

\begin{table}[h]
\centering
\begin{tabular}{c|c|c|c|c|c|c}
\hline\hline
 Field   &  Spin   & $SU(3)$ & $SU(2)_L$ & $U(1)_{\mathcal{Y}}$ & $Z_4$ & $B$ \\
\hline
$\psi_{L,R}$ & 1/2 & 3 & 1 & 2/3 & $i$ & 1/3 \\
$\eta$ &0 & 3 & 1 & 2/3 & $i$ & 1/3 \\
$N$ & 1/2 & 1 & 1 & 0 & $1$ & 0 \\
$\phi$ &0 & 1 & 1 & 0 & $i$ & 0 \\
\hline\hline
\end{tabular}
\caption{Representations of new particles in the Dirac Baryogenesis model. All the BSM particles are $SU(2)_L$ singlets, with electric charges $Q$ equal to the corresponding hypercharges.  }
\label{tab:particle}
\end{table}

\subsection{Mass spectrum and decay width}
\label{sec:mass}

\begin{figure}
\centering
\includegraphics[width=0.25\textwidth]{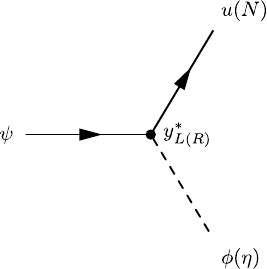}
\includegraphics[width=0.25\textwidth]{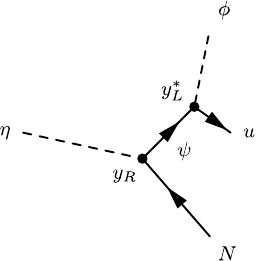}
	\caption{The Feynman diagrams for $\psi$ and $\eta$ decays.}
	\label{fig:feyndecay}
\end{figure}

While $\psi$ needs to be the heaviest particle in the model to realize baryogenesis from its decay, we consider the masses of $\phi$ and $N$ to be much lighter compared to $\eta$ for the sake of illustration. Therefore, we have the decay channels
\begin{eqnarray}
	\psi\to u\phi,\ \eta N, \ {\rm and}\   \eta\to N(\psi^* \to u\phi),
\end{eqnarray}
with the corresponding Feynman diagrams shown in Fig.~\ref{fig:feyndecay}. In this work, for simplicity, we only take the first generation up quark, though the proposed model applies equivalently to charm and top as well. 
As discussed in Eq.~(\ref{eq:Y}) of App.~\ref{app:Asym}, we take a universal Yukawa coupling strength (denoted by $y$) for simplification and a relative phase between two generations of $y_{Ri}$ to maximize the asymmetry. 
In addition, for representative purposes, we consider a benchmark choice
\begin{eqnarray}\label{eq:bench1}
M_{N}=20~\GeV, ~ M_{\phi}=62.5~\GeV,   
\end{eqnarray}
respecting to the mass hierarchy $M_{N,\phi}\ll M_{\eta}<M_{\psi}$. In fact, $M_{\phi} \sim 62.5$ GeV keeps the scalar singlet in the currently allowed ballpark for it to be a viable dark matter candidate~\cite{GAMBIT:2017gge}. 

\begin{figure}
	\centering
	\includegraphics[width=0.53\textwidth]{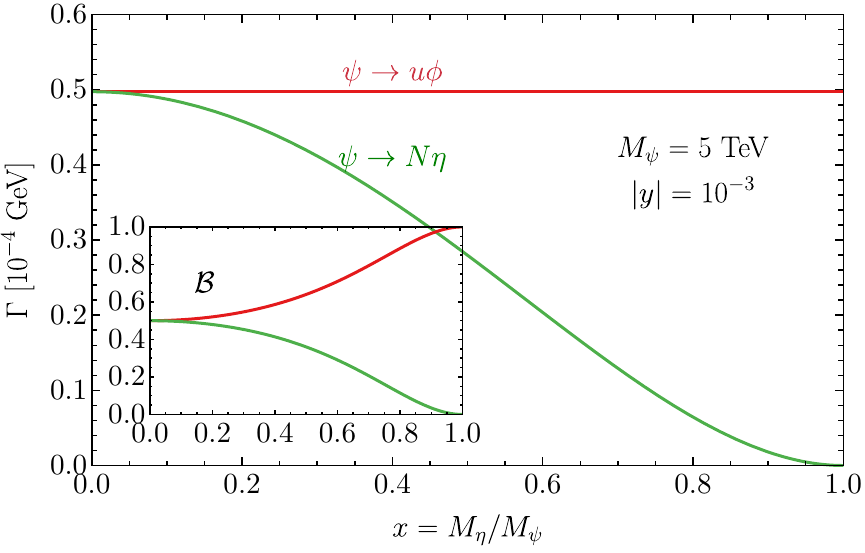}  
	\includegraphics[width=0.46\textwidth]{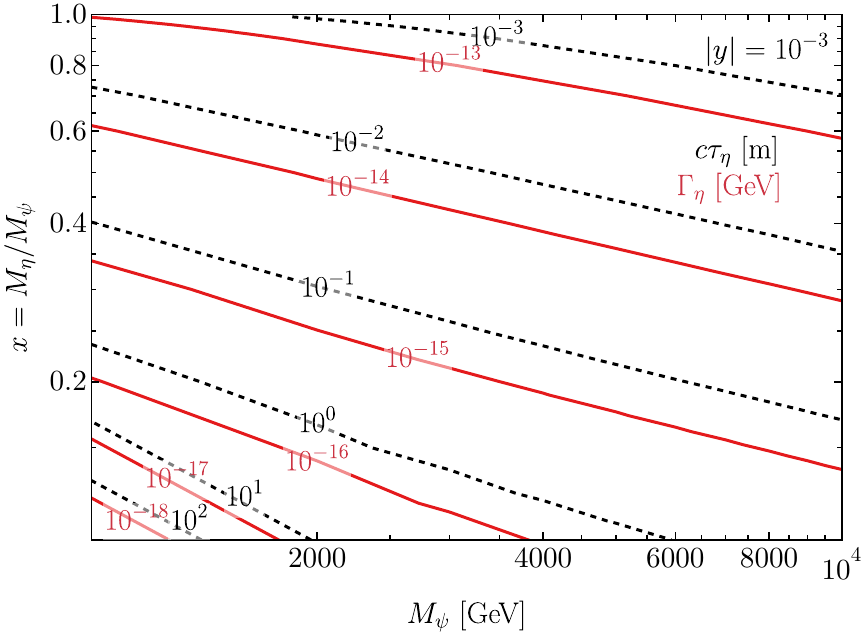}   
	\caption{Left: The tree-level decay widths and branching fractions (inner panel) for $\psi$, Right: for $\eta$ and the decay length $c\tau$ (dashed curves), with the benchmark parameter choices of Eqs.~(\ref{eq:bench1}) and (\ref{eq:Y}). }
	\label{fig:decay}
\end{figure}

The tree-level partial width of two-body decays of $\psi_i\to N\eta$ can be expressed as
\begin{equation}\label{eq:psidecay}
	\Gamma(\psi_i\to N\eta)=\frac{|y_{Ri}|^2}{32\pi} M_\psi\ \lambda^{1/2}(1,x^2_N,x^2_\eta) (1+x_N^2-x_{\eta}^2),
\end{equation} 
where $x_{N,\eta}=M_{N,\eta}/M_\psi$ and the K\"allen function is defined as 
$\lambda(a,b,c)=a^2+b^2+c^2-2ab-2bc-2ca$.  
The tree-level partial width of $\psi_i\to u\phi$ can be obtained similarly as
\begin{equation}
\Gamma(\psi_i\to u\phi)=\frac{|y_{Li}|^2}{32\pi}M_{\psi}
\end{equation}
with the condition $M_{u,\phi}\ll M_{\psi}$.
In Fig.~\ref{fig:decay} (left), we demonstrate the dependence of the partial decay widths and branching fractions (inner panel) of $\psi$ versus the mass ratio $x=M_{\eta}/M_{\psi}$ with the choice of $y=10^{-3}$ and $M_{\psi}=5~\TeV$. We see that in the massless limit $x\to0$, we have equal branching fractions for $\psi\to N\eta$ and $\psi\to u\phi$. In contrast, when $x\to1$, the $N\eta$ channel is asymptotically turned off, due to the phase space suppression. Overall, in the parameter space we are interested in, the $\psi$'s mainly decay promptly, which will guide our collider search strategies in Sec.~\ref{sec:collider}.

In comparison, the $\eta$ can only decay into 3 bodies via a virtual $\psi$ as in Fig.~\ref{fig:feyndecay} (right).
The total width depends on its mass $M_\eta$ as well as the propagator $M_\psi$, with a numerical demonstration in Fig.~\ref{fig:decay} (solid curves in the right panel). 
In the massless limit of the final states $M_{N,u,\phi}\to0$, the three-body width of $\eta$ can be simplified as
\begin{equation}
	\Gamma_\eta\simeq \frac{\left|\sum_{i}^{N_{\psi}}y_{Ri}y_{Li}\right|^2}{24(4\pi)^3}\frac{M_{\eta}^3}{M_\psi^2},
\end{equation}
when $M_{\eta}\ll M_\psi$. Keep in mind the massless three-body phase space factor $\Phi^{(3)}=M^2/4(4\pi)^3$.
Here, the index $i$ runs over the generation of the Dirac fermions with $N_{\psi}=2$, and we only consider the up quark.
When converting the decay width into the decay length as $c\tau_\eta$ in Fig.~\ref{fig:decay} (dashed curves in the right panel), we see that in the parameter space of our interest $\eta$ is long-lived in terms of the collider detector size. It can be searched through the displaced vertices as well as the colored tracks, as explored in Sec.~\ref{sec:collider}. However, we do not expect any astrophysical or cosmological effects, as the lifetime of $\eta$ is $\leq 10^{-6}$ s which remains much smaller than the BBN time scale, \emph{e.g.}, 1 s. 

\subsection{Loop induced asymmetry}
\label{sec:loop}

\begin{figure}
    \centering
    \includegraphics[width=0.49\textwidth]{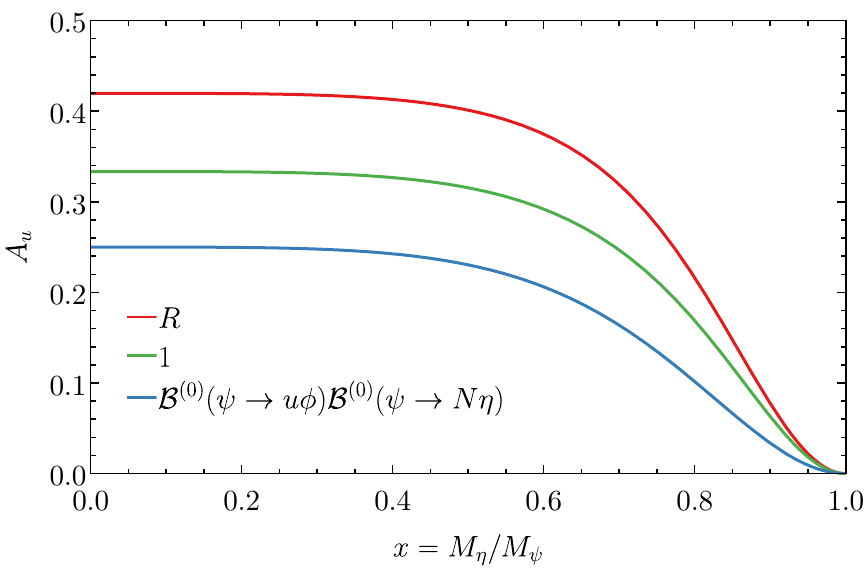}
    \includegraphics[width=0.49\textwidth]{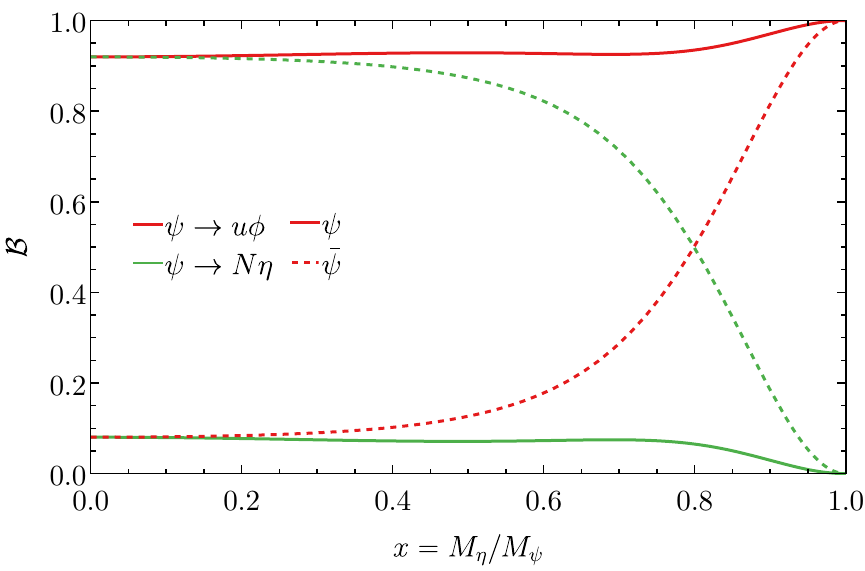}    
    \caption{Left: The dependence of the up quark asymmetry $A_u$ in the first recursive (1) and resummed (R) solutions, with respect to the zero-th order approximation $\mathcal{B}^{(0)}(\psi \to u \phi)\mathcal{B}^{(0)}(\psi \to N\eta)$.
    Right: The dependence of the branching fractions of $\psi$ and $\bar{\psi}$ on the mass ratio $x=M_{\eta}/M_\psi$ in the resummation calculation.}
    \label{fig:AuBR}  
\end{figure}

As explained in detail in App.~\ref{app:Asym}, an asymmetry in the SM quark sector can be generated from the interference of tree and one-loop self-energy diagrams shown in Fig.~\ref{fig:LEPTO}. Two different copies of vector-like quark $\psi$ give rise to the self-energy correction responsible for generating the CP asymmetry.
We can define the baryonic asymmetry of the $u$ quark in the SM sector as
\begin{equation}
	\label{eq:Au}
	\begin{aligned}
		A_u &= \frac{\Gamma(\psi\rightarrow u\phi)-\Gamma(\bar{\psi}\to\bar{u}\bar{\phi})}{\Gamma(\psi\to u \phi)+\Gamma(\bar{\psi}\to\bar{u}\bar{\phi}) + \Gamma(\psi\to N \eta)+\Gamma(\bar{\psi}\to N\bar{\eta})}.
	\end{aligned}    
\end{equation}
To the leading order approximation, we can relate $A_u$ to the baryon-antibaryon asymmetry $\delta_u$ defined in Eq.~(\ref{eq:delta}) as 
\begin{equation}
A_u^{(0)}=\calB^{(0)}(\psi\to u\phi)\delta_u ,
\end{equation}
where $\calB^{(0)}$ is the decay branching fraction. Assuming $\Delta M_{\psi}\sim \Gamma/2$, the maximal value of the CP asymmetry parameter $\delta_u$ can be found as
\begin{equation}
\delta_{\max}\sim\frac{\Im[y_{R1}^*y_{Rj}y_{Lj}y_{L1}^*]}{|y_{R1}|^2|y_{L1}|^2}\calB^{(0)}(\psi_1\to N\eta),
\end{equation}
with details to be found in App.~\ref{app:Asym}. Additionally, we also consider the Yukawa universality relation in Eq.~(\ref{eq:Y}) to maximize the $A_u$, similar to the resonant leptogenesis~\cite{Pilaftsis:1997jf}. As a consequence, the $u$-quark asymmetry can be approximated as
\begin{equation}
	A_u^{(0)}\sim \mathcal{B}^{(0)}(\psi \to u \phi)\mathcal{B}^{(0)}(\psi \to N\eta),
\end{equation}
with $\mathcal{B}^{(0)}$ denoting the leading order branching fractions. As before, with the mass hierarchy $M_{N,\phi}\ll M_\eta<M_\psi$, we can obtain a simple dependence of the zero-th $A_u$ on the mass ratio $x=M_\eta/M_\psi$, shown as the blue line in Fig.~\ref{fig:AuBR} (left). In the small mass limit $M_{\eta}\ll M_\psi$, the asymmetry approaches a constant 
\begin{equation}
A_u^{(0)}\sim 1/4.
\end{equation}

However, as demonstrated in App.~\ref{app:loop}, besides the asymmetry $\delta_u$, the partial widths for decays $\psi\to u\phi/N\eta$ and $\bar{\psi}\to\bar{u}\bar\phi/N\bar{\eta}$ face large loop corrections. Therefore, we have to consider corrections to the denominator in Eq.~(\ref{eq:Au}) (left) as well. With the recursion and resummation methods developed in App.~\ref{app:loop}, we show the first recursive and resummed results for $A_u$ as the green and red lines in Fig.~\ref{fig:AuBR} (left), respectively. We see that even without capturing the exact numerics, the zero-th order solution $A_u^{(0)}$ gives a correct shape for $A_u$ already. With the first recursion ($A_u^{(1)}$) and resummation ($A_u^{(R)}$), the $A_u$ will be corrected up to 
\begin{equation} 
A_u^{(1)}=1/3, ~\textrm{or}~A_u^{(R)}=0.420,
\end{equation}
respectively, when $M_\eta\ll M_\psi$. Meanwhile, we also present the resummed branching fractions of $\psi$ and $\bar{\psi}$ in Fig.~\ref{fig:AuBR} (right). We see that the $\psi$ dominantly decays to $u\phi$ while its anti-particle $\bar{\psi}$ mainly decays into $N$ and $\bar{\eta}$, with 
\begin{equation}
    \br(\psi\to u\phi) \simeq \br(\bar\psi\to N\bar{\eta}) \simeq 1,
\end{equation}
when $x=M_{\eta}/M_{\psi}\ll1$. 

In the rest of this work, except specifying explicitly, we will adopt the resummation method in both baryogenesis and collider phenomenology in Sec.~\ref{sec:cosmo} and \ref{sec:collider}, respectively.

\section{Generation of baryon asymmetry}
\label{sec:cosmo}

\begin{figure*}[!h]
  \centering
\includegraphics[width=0.24\textwidth]{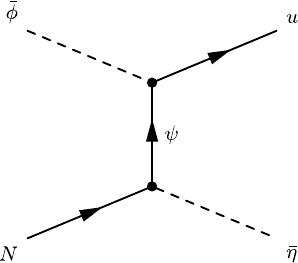}
\includegraphics[width=0.24\textwidth]{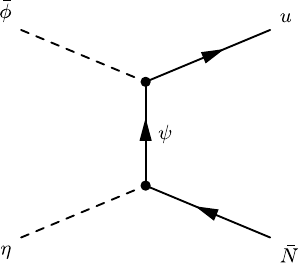}
\includegraphics[width=0.24\textwidth]{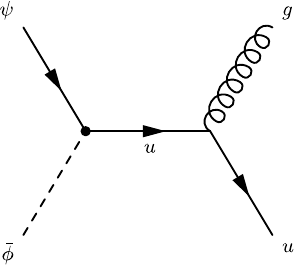}
\includegraphics[width=0.24\textwidth]{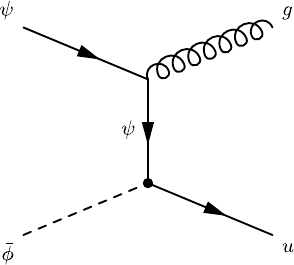}
\caption{Representative washout processes, with the $\eta$ decay in Fig.~\ref{fig:decay}~(right).}
\label{fig:WO}
\end{figure*}

\begin{figure}[!h]
    \centering
\includegraphics[width=0.24\textwidth]{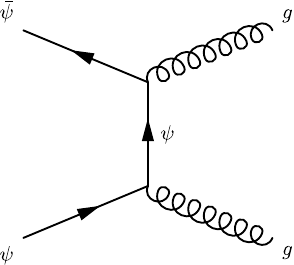}
\includegraphics[width=0.24\textwidth]{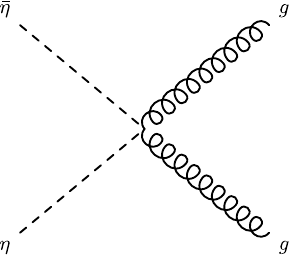}    
    \caption{Dominant processes keeping $\psi, \eta$ in equilibrium with the bath.}
    \label{fig:ANN}
\end{figure}

In addition to the decay processes responsible for generating the asymmetries and the corresponding inverse decays responsible for washout, there are two other types of processes: (i) washout scattering processes which reduce the asymmetries generated in each sector or reduce the baryon asymmetry in SM quark sector by redistribution of asymmetries in the two sectors, (ii) processes which keep the mother particle in equilibrium for longer epoch delaying the asymmetry generation. The dominant processes belonging to category (i), (ii) are shown in Fig. \ref{fig:WO} and Fig. \ref{fig:ANN} respectively.

\begin{figure}
\centering
\includegraphics[scale=0.45]{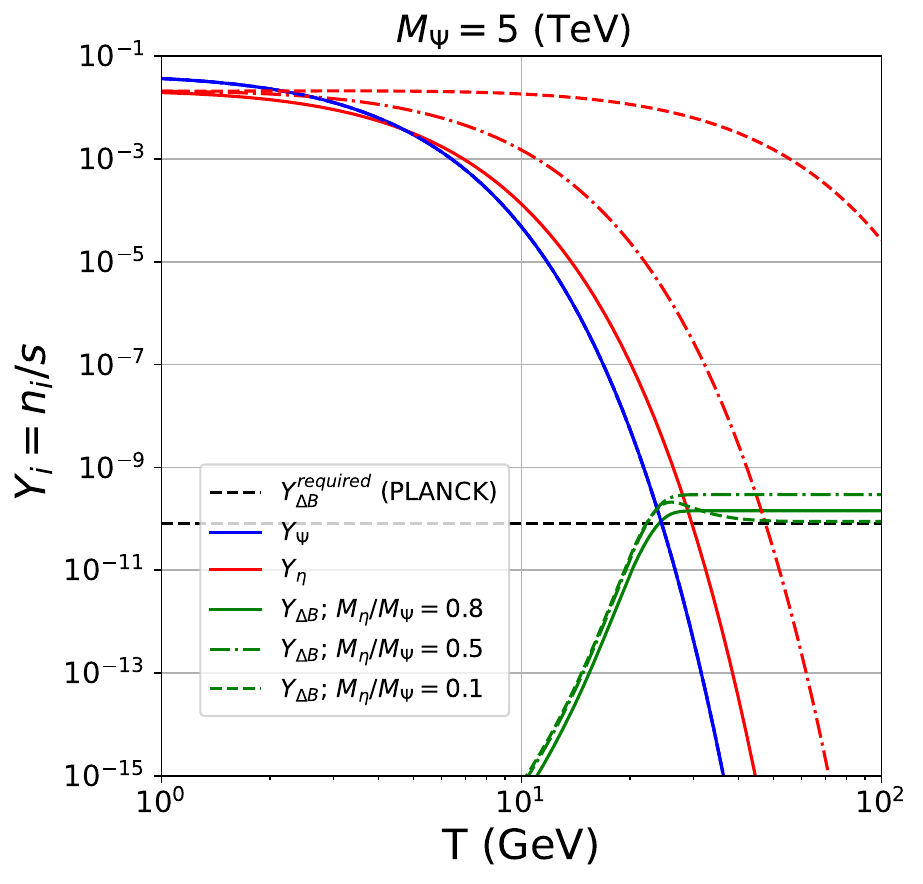}
\includegraphics[width=0.54\textwidth]{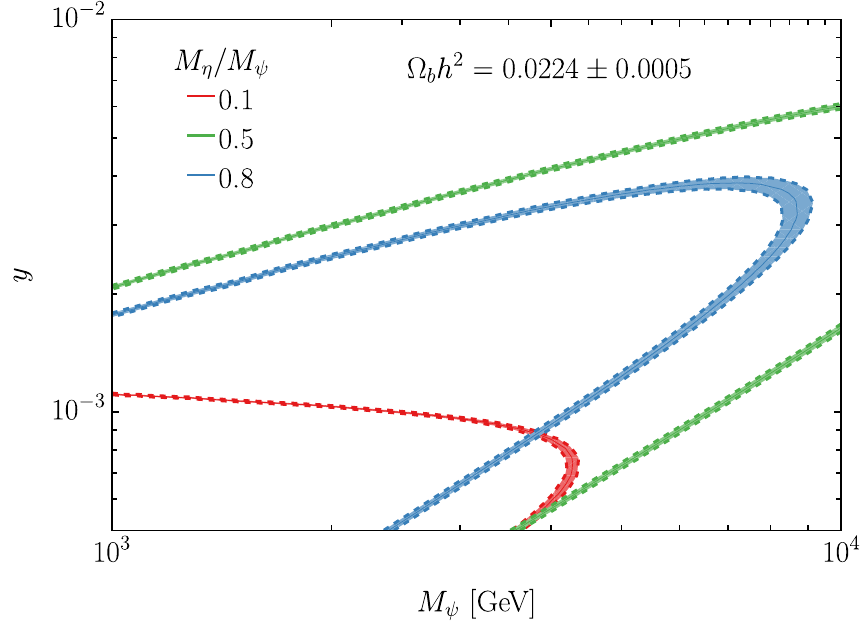}
\caption{
Left:  Evolution of comoving abundances $Y_i=n_i/s$ of baryon asymmetry $\Delta B$, $\psi$ and $\eta$ with fixed $M_\psi = 5$ TeV, $y=10^{-3}$ in Eq.~(\ref{eq:Y}) and $M_\eta/M_\psi =0.1,~0.5,$ and 0.8, respectively.
Right: The parameter space of $\psi$ mass and Yukawa coupling $(M_\psi,y)$ consistent with the observed baryon asymmetry $\Omega_b h^2 = 0.0224\pm0.0005$ within $5\sigma$~\cite{Planck:2018vyg} for fixed mass ratios $M_\eta/M_\psi$. }
\label{fig:BG}
\end{figure}

Including all these relevant processes into account, the Boltzmann equations required for calculating the baryon asymmetry in SM sector are given by
\begin{equation}
\begin{aligned}
\frac{dY_\psi}{dz} &= -\frac{1}{Hz}\left(\langle \Gamma_{\rm total} \rangle(Y_\psi - Y^{\rm eq}_\psi) + s\langle \sigma_{\psi \psi \rightarrow gg} v \rangle \left(Y^2_\psi - (Y^{\rm eq}_\psi)^2\right)\right),  \\
\frac{dY_\eta}{dz} &= -\frac{1}{Hz}\left(\langle \Gamma(\eta \rightarrow u \phi N)) \rangle(Y_\eta - Y^{\rm eq}_\eta) + s\langle \sigma(\eta \eta\rightarrow gg) v \rangle \left(Y^2_\eta - (Y^{\rm eq}_\eta)^2\right)\right),  \\
\frac{dY_{\Delta B}}{dz} &= \frac{1}{Hz}\bigg [A_u\langle \Gamma(\psi\rightarrow u \phi) \rangle (Y_\psi - Y^{\rm eq}_\psi)
- Y_{\Delta B} \frac{Y^{\rm eq}_\psi}{Y^{\rm eq}_{\Delta B}}\left ( \langle\Gamma(\psi\rightarrow u \phi)\rangle + s\langle \sigma(\psi \phi \rightarrow g u) v\rangle \right ) \\ 
&- Y_{\Delta B}\frac{Y^{\rm eq}_\eta}{Y^{\rm eq}_{\Delta B}}\left(\langle\Gamma(\eta\rightarrow u \phi N)\rangle + s\langle \sigma(\eta \phi \rightarrow N u)v\rangle + s\langle \sigma(\eta N \rightarrow \phi u)v\rangle\right)\bigg ].
\end{aligned}  
\label{eq:BE}
\end{equation}
Here $Y_i=n_i/s$ denotes the comoving number density and $z=M_\psi/T$ is a dimensionless variable. Here, $\langle \Gamma_x \rangle = \Gamma_x K_1 (M_x/T)/K_2(M_x/T)$ is the thermal averaged decay width of particle $x$ with $K_i$'s being modified Bessel functions of i-th order. $\langle{\sigma v}\rangle$ is the thermally averaged annihilation cross-section defined by \cite{Gondolo:1990dk}
\begin{equation}
    \langle \sigma v \rangle_{ij \rightarrow kl} = \frac{1}{8Tm^2_i m^2_j K_2 (z_i) K_2 (z_j)} \int^{\infty}_{(m_i+m_j)^2} ds \frac{\lambda (s, m^2_i, m^2_j)}{\sqrt{s}} K_1 (\sqrt{s}/T) \sigma
\end{equation}
with $z_i=m_i/T$ and $\lambda (s, m^2_i, m^2_j)=[s-(m_i+m_j)^2][s-(m_i-m_j)^2]$. For colored particles, we consider the dominant annihilation channel only, namely, the annihilation into gluons. However, for coannihilations, we consider the dominant processes responsible for washout. The first and the last Boltzmann equations mentioned above are the standard Boltzmann equations for the mother particle and baryon asymmetry respectively. We also solve the Boltzmann equation for $\eta$ which need not be in equilibrium throughout and can induce large washout via three-body decay or scattering into SM quarks. Since $\eta$ and SM quarks store equal and opposite CP asymmetries, the decay of $\eta$ to SM quarks can lead to washout of asymmetry stored in the latter. For a few chosen benchmark points shown in Table \ref{tab:benchmark}, we solve these coupled Boltzmann equations numerically. The evolution of comoving densities are shown in Fig. \ref{fig:BG} (left). If we decrease the mass of $\eta$ while keeping $M_\psi$ fixed, initially we see a rise in asymmetry due to enhanced CP asymmetry. However, if we keep lowering the mass of $\eta$ further, at some point, the washout induced by $\eta$ will be too strong and lead to a fall in asymmetry. This is clearly seen in the evolution of baryon asymmetry for the three benchmark points of Table \ref{tab:benchmark}. We also perform a numerical scan over the parameter space consistent with the observed baryon asymmetry. The resulting parameter space is shown in Fig. \ref{fig:BG} (right) in the plane of Yukawa coupling $y$ versus $M_\psi$ where contours of fixed mass ratio $M_\eta/M_\psi$ are consistent with the observed baryon asymmetry. As we decrease $M_\eta$, the parameter space initially widens up due to enhanced CP asymmetry. However, with further decrease in $M_\eta$, the washout from late decay of $\eta$ dominates, leading to shrinking of the parameter space, a feature also noticed in the evolution of asymmetries shown in Fig. \ref{fig:BG} (left). Most of the baryon asymmetry contours in Fig. \ref{fig:BG} (right) show two different values of Yukawa coupling due to the interplay of production and washout. While smaller $y_{L,R}$ slows down the production, larger $y_{L,R}$ comes with enhanced washout, though they do not affect the CP asymmetry in the resonant regime.

While we are not solving the Boltzmann equation for DM $\phi$, it can freeze-out at a temperature much below the scale of baryogenesis due to the chosen mass hierarchy $M_{\phi}\ll M_\psi$, in a way similar to thermal scalar singlet dark matter \cite{GAMBIT:2017gge}. In the simplest setup with unbroken $Z_4$ symmetry, constraints on DM relic \cite{Planck:2018vyg} and direct-detection rate \cite{LZ:2022lsv} leave only a tiny window of parameter space near the SM Higgs resonance region \cite{GAMBIT:2017gge, DiMauro:2023tho}.

\begin{table}[]
    \centering
    \begin{tabular}{c|c|c|c|c|c|c}
    \hline
         &$y$ & $M_\psi$~[TeV] & $M_\eta$~[TeV] & $M_N$~[GeV] & $M_\phi$~[GeV] & $A_u^{(R)}$ \\
        \hline
        \hline
         BP1 & $10^{-3}$ &    5   &   0.5    &  20   &   62.5  & 0.420   \\
         BP2 & $10^{-3}$ &    5   &   2.5    &  20   &   62.5  & 0.401   \\
         BP3 & $10^{-3}$ &    5   &   4.0    &  20   &   62.5  & 0.216   \\
         \hline
     \end{tabular}
    \caption{Benchmark values of parameters used for the evolutions shown in Fig. \ref{fig:BG} (left). The Yukawa coupling $y$ is chosen to maximize the phase as Eq.~(\ref{eq:Y}). }
    \label{tab:benchmark}
\end{table}
   
\section{Collider Signals}
\label{sec:collider}

\begin{figure}
	\centering
	\includegraphics[width=0.24\textwidth]{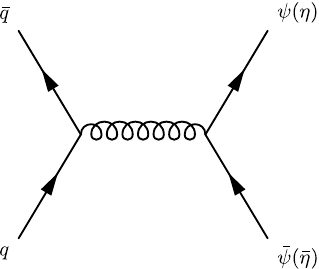}
	\includegraphics[width=0.24\textwidth]{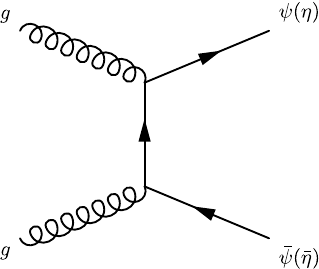}
	\includegraphics[width=0.24\textwidth]{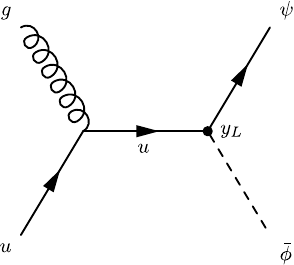}
	\includegraphics[width=0.24\textwidth]{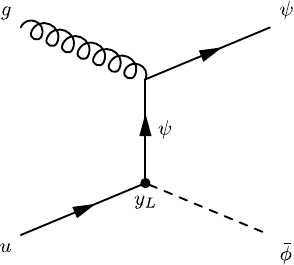}           
	\caption{Representative Feynman diagrams for the $\psi$ and $\eta$ production at hadron colliders.}
	\label{fig:feynLHC}
\end{figure}

With the benchmark parameters considered in our model in Table \ref{tab:benchmark}, it is conceivable to search for the TeV scale new particles at the current and future high-energy colliders. 
The LHC and its luminosity upgrade (HL-LHC) will lead the new physics searches at the energy frontier. Future multi-TeV lepton colliders will open a new territory for the discovery potential. Therefore, we will consider both hadron and lepton colliders, such as the LHC and a high-energy muon collider.

\begin{figure}
    \centering
    \includegraphics[width=0.49\textwidth]{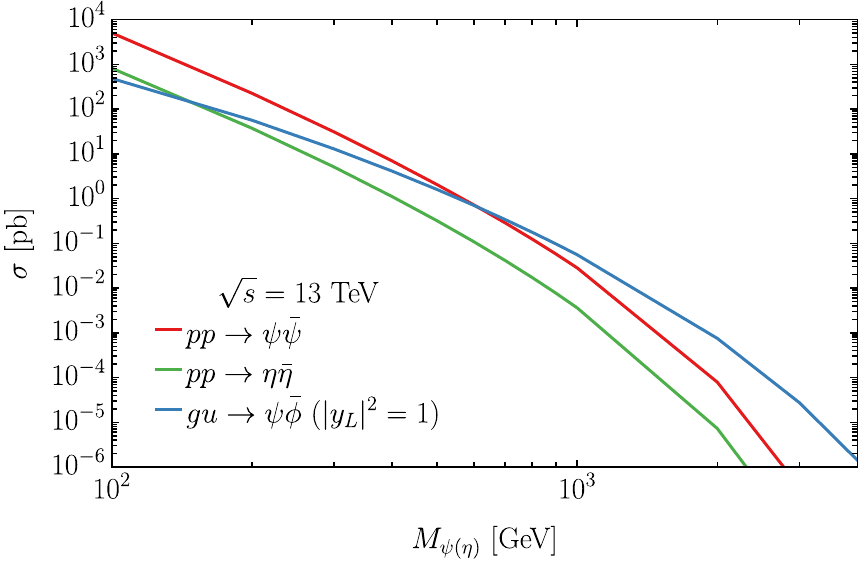}
    \includegraphics[width=0.49\textwidth]{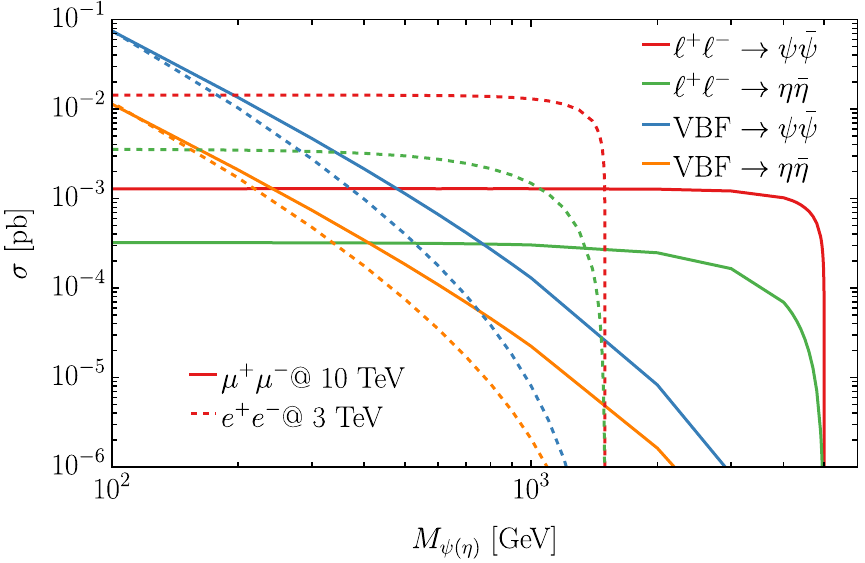}    
    \caption{Left: The production cross-sections of a pair of $\psi$ and $\eta$, respectively, at the 13 TeV LHC (left). Right: Future lepton colliders with $\sqrt{s}=3$ TeV (dashed curves) and 10 TeV (solid curves). }
    \label{fig:xsLHC}
\end{figure}

At hadron colliders, the colored particles $\psi$ and $\eta$ can be copiously produced in pairs through QCD processes, as shown in the left two diagrams of Fig.~\ref{fig:feynLHC}.
Since $\psi$ couples to light quarks (\emph{e.g.}, $u$) via Yukawa interaction, we have an additional single production mechanism, as shown in the right two diagrams of Fig.~\ref{fig:feynLHC}. The dependence of the pair and single production at the 13 TeV LHC on the particle mass $M_{\psi(\eta)}$ is shown in Fig.~\ref{fig:xsLHC} (left). 
The calculation is done at the leading order with \texttt{MadGraph}~\cite{Alwall:2014hca,Frederix:2018nkq} and \texttt{Whizard}~\cite{Kilian:2007gr,Moretti:2001zz,Christensen:2010wz}, interfacing with the \texttt{UFO}~\cite{Degrande:2011ua} model file generated with \texttt{FeynRules}~\cite{Alloul:2013bka} and \texttt{SARAH}~\cite{Staub:2013tta}, and parton distribution functions (PDFs) from CT18 NNLO~\cite{Hou:2019efy}. 
For the single $\psi$ production, in Fig.~\ref{fig:xsLHC} we take the $|y_L|^2=1$ as a benchmark, while other values can be rescaled with $\sigma\propto|y_L|^2$ correspondingly. 

\begin{figure}
	\centering
	\includegraphics[width=0.24\textwidth]{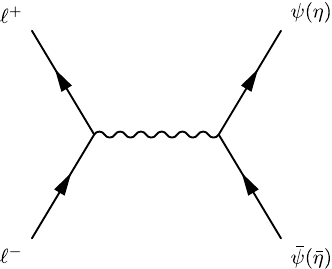}\quad
	\includegraphics[width=0.28\textwidth]{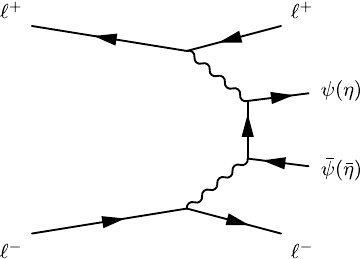}    
	\caption{Representative  Feynman diagrams for pair production of $\psi$ and $\eta$ at a high-energy lepton collider via (left) the direct annihilation and (right) gauge boson fusion.}
	\label{fig:feynMuC}
\end{figure}

Since both $\psi$ and $\eta$ carry electric charges as well, they can be produced through electroweak (EW) processes, such as annihilation or fusion, as shown in Fig.~\ref{fig:feynMuC}. Similar processes can happen at hadron colliders with initial quarks as well, while the cross-section is suppressed with respect to the QCD one from Fig.~\ref{fig:feynLHC}. At lepton colliders, the EW scatterings dominate, while the QCD ones are suppressed due to the low parton luminosity of quark and gluon~\cite{Han:2021kes}.
In Fig.~\ref{fig:xsLHC} (right), we show the pair production rates of $\psi$ and $\eta$ at electron and muon colliders, with collision energy as $\sqrt{s}=3$ and 10 TeV, respectively, corresponding to the setup of CLIC~\cite{CLICdp:2018cto} and future muon collider~\cite{Accettura:2023ked,Bartosik:2020xwr}. 
The fusion is calculated with photon-initiated $2\to2$ process with equivalent photon approximation~\cite{Fermi:1924tc,vonWeizsacker:1934nji,Williams:1935dka}.  The numerical difference between the equivalent photon approximation and the full $2\to4$ process of Fig.~\ref{fig:feynMuC} (right) is found to be generally less than 10\%, with an exception around $2M_{\psi(\eta)}\sim\sqrt{s}$ where the fusion cross-section is completely negligible. As we expect, the fusion process only dominates at a low $\psi/\eta$ mass near the threshold, while the annihilation takes over especially in the parameter space of interest when $M_{\psi(\eta)}>1~\TeV$. 

As both $\psi$ and $\eta$ finally decay to a quark and some invisible particles, the signals from the prompt decay of $\psi$ and $\eta$ (pair) are a mono-jet (di-jet) with a large missing energy.
In the parameter space we consider, $\eta$ decays to 3-body final states, so its width is small and it can be long-lived or nearly stable.  In this case the displaced vertex or $R$-hadron tracks~\cite{Buckley:2010fj,ATLAS:2019gqq,Ghosh:2017vhe} are background-free signals of $\eta$.

In our model, to generate the asymmetry as discussed in Sec.~\ref{sec:loop}, we need at least two $\psi$'s.
The production cross-section can be written in the narrow width approximation form as
\begin{equation}
\begin{aligned}
&\sigma(pp\to \bar\phi \psi \to \bar{\phi} X)=\sigma(pp\to \bar\phi \psi)\br(\psi \to  X)\\
&\sigma(pp\to \psi\bar{\psi} \to X\bar{X})=\sigma(pp\to \bar \psi \psi) \br(\psi \to X)\br(\bar{\psi}\to\bar{X}),
\end{aligned}
\end{equation} 
where $X$ denotes any decay final state of $\psi_i$.
As demonstrated in Sec.~\ref{sec:loop} and more details in App.~\ref{app:Asym}, owing to the intrinsic CP asymmetry, the decay branching fractions of the particle and anti-particle are different, $\br(\psi \to X)\neq \br(\bar{\psi}\to\bar{X})$.
Due to the same decay modes of $\psi_1$ and $\psi_2$, the production cross-section doubles when we consider both contributions, assuming the interference between $\psi_1$ and $\psi_2$ to be negligible. However, in the nearly degenerate case between $\psi_1$ and $\psi_2$, \emph{i.e.}, $\Delta M_\psi \equiv M_{\psi_1}-M_{\psi_2} = \alpha\Gamma_\psi$ with $\alpha\lesssim1$, the $\psi_1$ and $\psi_2$ resonances overlap with each other so that interference effects must be taken into account.
With the benchmark coupling choice in Eq.~\eqref{eq:Y} and considering the interference effect of the tree-level amplitudes (see App.~\ref{app:BreitWigner} for details), the total production cross-sections of $\psi_i$'s are 
\begin{align}\label{eq:singlepsi12interfer}
\sum_{i=1,2}\sigma(pp\to \bar\phi \psi_i(\to u\phi)) &= 2\left(1+\frac{1}{1+\alpha^2}\right)\sigma(pp\to \bar\phi \psi(\to u\phi)), \\
\label{eq:psipair12interfer}
\sum_{i=1,2}\sigma(pp\to \psi_i (\to u \phi) \bar \psi_i (\to N\bar\eta)) &= 2\left(1+\frac{2\alpha}{(1+\alpha^2)^2}\right)\sigma(pp\to \psi(\to u \phi) \bar \psi (\to N\bar\eta)).
\end{align}
In comparison with one generation of $\psi$, the total cross-section of $\psi_i$'s (pair) production is obtained with a scale factor.
The scale factor is 2 when $\Delta M_\psi \gg \Gamma_\psi$ ($\alpha\gg 1$) such that there is no interference. 
In this work, we take $\alpha=1/2$ to maximize the asymmetry; see App.~\ref{app:Asym} for details.

\subsection{Constraints from current LHC data}
\subsubsection{Mono-jet signal from the prompt decay $\psi\to u\phi$}
\label{sec:monojet}

\begin{figure}
    \centering
    \includegraphics[width=0.47\textwidth]{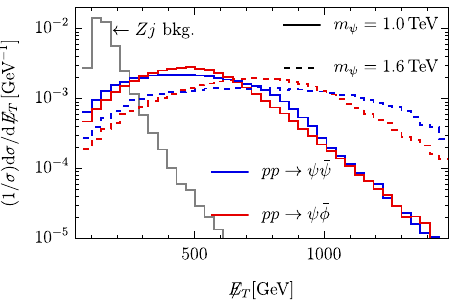}    ~~
    \includegraphics[width=0.47\textwidth]{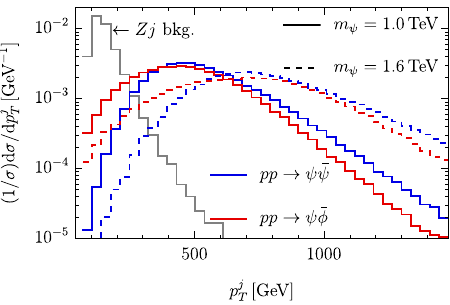}    
    \caption{The distribution of missing transverse energy $\slashed{E}_T$ (left) and leading jet $p_T$ (right) from the SM backgrounds (gray)~\cite{ATLAS:2017bfj}, $\psi$-pair production (blue) and single-$\psi$  production (red) at the 13 TeV LHC.}
    \label{fig:1jmet}
\end{figure}

Recall that the particle $\psi$ mainly decays through $u\phi$, while its anitparticle $\bar{\psi}$ decays into $N\bar{\eta}$, as shown in Fig.~\ref{fig:AuBR} (right). As a consequence, both the pair $pp\to \psi\bar \psi$ and single $pp\to \psi\phi$ production processes will lead to a mono-jet signal with large missing energy. The most prevalent SM mono-jet background is $Z(\to \nu\bar \nu)$+jets~\cite{Schramm2017,ATLAS:2017bfj,ATLAS:2021kxv}.  In contrast to the background, the transverse momenta of the $\psi$ decay products are substantially higher $\sim M_\psi / 2$ due to the Jacobian peak in the two-body decay, as shown in  Fig.~\ref{fig:1jmet}.
Therefore, the $\psi\to u\phi$ signal and the SM backgrounds can be effectively distinguished by analyzing the distribution of the transverse momentum of the leading jet $p_T^j$ and the missing transverse energy $\slashed{E}_T$. 

We generate the signal event at parton level with \texttt{MadGraph}~\cite{Alwall:2014hca,Frederix:2018nkq}, and the parton level events are passed to \texttt{Pythia8}~\cite{Bierlich:2022pfr} for parton shower and hadronization, where $\eta$ is treated as a missing particle.\footnote{In fact, $\eta$ would hadronize to an $R$-hadron in a way similar to the supersymmetric (SUSY) particle stop $\tilde{t}$, which can leave a colored track in the calorimeter.  Here, we do not take the $R$-hadron signal into account, as $\eta$ is mainly produced from $pp\to \eta \bar \eta$ processes instead of $\bar \psi$ decay.}
The $\met$ and $p_T^j$ distributions of the signals and the leading background are shown in Fig.~\ref{fig:1jmet}. To discriminate the highly energetic mono-jet signal against the backgrounds, we first adopt the acceptance cuts
\begin{equation}
    p_T^j({\rm leading})>250~{\GeV},\ |\eta_j|<2.4;\quad \met > 250~{\GeV}. 
\end{equation}
where $p_T^j$ and $\eta_j$ are the transverse momentum and the rapidity of the leading jet.
We allow extra jet activities in the events with a maximum of four jets with $p_T>$ 30 GeV and rapidity $|\eta|< 2.8$. In addition, separation in the azimuthal angle of $\Delta\varphi(\vec p_j,\vec p_{\rm miss})> 0.4$ between the missing transverse momentum direction and each selected jet is required to reduce the multijet background.

\begin{figure}
    \centering
    \includegraphics[width=0.49\textwidth]{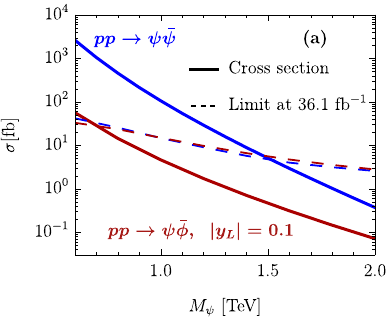}
    \includegraphics[width=0.49\textwidth]{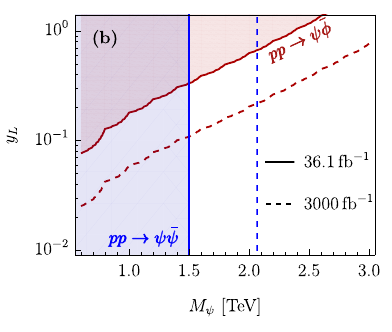}
    \caption{Left: Summary of expected upper limits at 95\% CL on the single-$\psi$ or $\psi$-pair production cross-section from mono-jet signals at the 13 TeV LHC. Right: The parameter space $(M_\psi,y_L)$ probed by current mono-jet searches (shaded region) and projections with luminosity scaled to $3000~\text{fb}^{-1}$ (dashed line).  Here, we adopt the resummed branching fraction in App.~\ref{app:Asym} assuming that  $M_\eta \ll M_\psi$.  The interference between $\psi_1$ and $\psi_2$ are calculated with Eq.~\eqref{eq:singlepsi12interfer} and~\eqref{eq:psipair12interfer}. }
    \label{fig:Prompt-Constrain}
\end{figure}

To optimize the constraint for different $M_\psi$, the $\met/\rm GeV$ of the pre-selected events is divided into 10 bins: [250, 300, 350, 400, 500, 600, 700, 800, 900, 1000] and above.
With the SM mono-jet background under the same selection cut~\cite{ATLAS:2017bfj}, we calculate the constraint on $\psi\bar\psi$ and $\psi\bar\phi$ production with $1j+\met$ final states using the likelihood ratio for the binned distribution in $\met$. The current constraints and future projections of our model parameter space from mono-jet signals are summarized in Fig.~\ref{fig:Prompt-Constrain}. 
The upper limits of the total cross-sections of $\psi\bar \psi$ and $\psi\bar\phi$ processes are similar because they produce similar mono-jet and $\slashed{E}_T$ distributions. The limit on the $pp\to \psi\bar\phi$ cross-section can be translated into a constraint on the parameter space $(M_\psi,y_L)$ as shown in Fig.~\ref{fig:Prompt-Constrain} (right),  while the $\psi$-pair production only provides a lower limit of $M_\psi$ in the parameter space we consider.\footnote{ In addition to the QCD process shown in Fig.~\ref{fig:feynLHC}, $\psi$-pair also can be produced through Yukawa coupling $y_L$ with a $\phi$ mediated $t$-channel process, but the overall contribution is still negligible when $|y_L|<1$. }

\subsubsection{The displaced vertex and colored track for $\eta$}
\label{sec:DV}
While resonantly enhanced CP asymmetry in the TeV-scale baryogenesis framework does not depend upon Yukawa couplings, keeping the latter small can keep the washout processes under control. Such small Yukawa couplings, on the other hand, can lead to a long-lived $\eta$ with displaced vertex (DV) being one of its characteristic signatures.  With a simplified spherical detector with radius $l_1$, a DV can be reconstructed when a $\eta$ decays after it travels for a minimum length $l_0$ but before reaching the detector. Then the number of events with at least one DV can be obtained with
\begin{equation}
N_{\rm DV}=\calL\sigma(M_\eta) \times (2P_{DV}-P_{\rm DV}^2),
\end{equation}
where the probability that a $\eta$ particle leaves a DV within the detector is defined as
\begin{equation}\label{eq:pDVsperical}
P_{\rm DV}=\int_{l_0}^{l_1}\frac{\dd r}{d}\exp(-\frac{r}{d})=e^{-l_0/d}-e^{-l_1/d}.
\end{equation}
Here, $d=\gamma\beta \tau_\eta=p_\eta/(M_\eta\Gamma_\eta)$, $\gamma=E_\eta/M_\eta$, $\beta=p_\eta/E_\eta$ are the corresponding decay length, boost factor and velocity factor respectively. 
In this work, we take the luminosity $\mathcal{L}=36.1~\textrm{fb}^{-1}$ from the CMS analysis~\cite{CMS:2021tkn}, and also project for the HL-LHC one with $\mathcal{L}=3000~\textrm{fb}^{-1}$~\cite{ZurbanoFernandez:2020cco}. Meanwhile, we can also observe two displaced vertices, with the event number estimated as
\begin{eqnarray}
N_{2\rm DV}=\calL\sigma(M_\eta)P_{\rm DV}^2.
\end{eqnarray}

\begin{figure}
\centering
\includegraphics[width=0.49\textwidth]{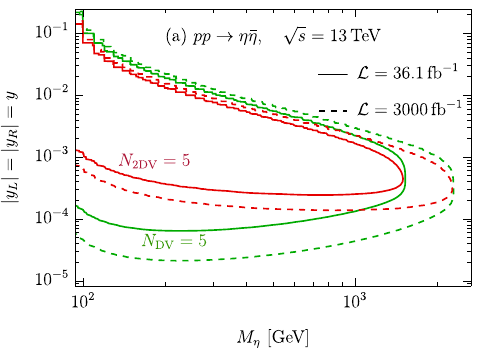}
\includegraphics[width=0.49\textwidth]{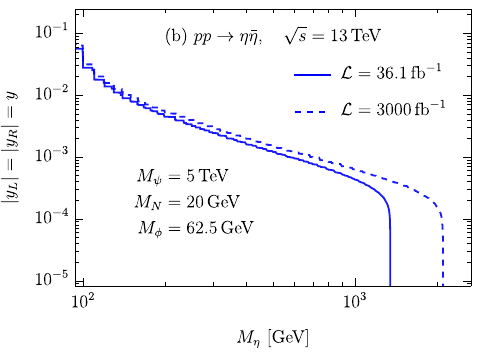}
\caption{
\label{fig:dv_and_Rhadron}
Left: The expected contours of $N_{\rm DV}=5$ and $N_{\rm 2DV}=5$  displaced vertex events for $\eta$ production.
Right: Constraint from $R$-hadron tracks on the $(M_\eta,y)$ parameter plane from the LHC data of 36.1 fb$^{-1}$ (solid) and scaled to $3000~\mathrm{fb}^{-1}$ (dashed).  Both with the benchmark parameter choice $M_{\psi}=5~{\rm TeV}$, $M_N=20{~\rm GeV}$ and $M_{\phi}=62.5{~\rm GeV}$.  The lifetime of $\eta$ is calculated with only the first generation of the up-type quark.}
\end{figure}

In a more realistic scenario with a cylinder detector of radius $r$ and height $2z$~\cite{ATLAS:2020xyo}, the distance between the displaced vertex and beamline in the transverse direction is required to be larger than $d_T$.
The probability in Eq.~\eqref{eq:pDVsperical} is then refined as
\begin{equation}
P_{\rm DV}=\left(e^{-d_T/(d\sin\theta)}-
\begin{cases}
e^{-r/(d\sin\theta)} & |\tan\theta|>r/z\\
e^{-z/(d\cos\theta)} & |\tan\theta|<r/z
\end{cases}\right).
\end{equation}
The experimental parameters for the muon tracking are~\cite{ATLAS:2020xyo}
\begin{equation}
|d_0|>2~\textrm{mm}, |d_0|<300~\textrm{mm}, ~|z_0|<300(1500)~\textrm{mm}.
\end{equation}
In Fig.~\ref{fig:dv_and_Rhadron} (left), we show the $N_{\rm DV}=5$ displaced vertex events contour for $\eta$ in the $(M_\eta,y)$ plane and we require $M_{\eta}>M_N+M_\phi$ to open the three-body decay channel $\eta\to N_L u \bar \phi$.  

When the decay length of $\eta$ becomes longer than the detector size, the scalar $\eta$ does not decay inside the detectors, and there is also no DV signal. In this case, the colored stable particle $\eta$ would hadronize, forming so-called $R$-hadrons~\cite{Farrar:1978xj}, and leave tracks inside the hadronic calorimeter.  
The ATLAS muon spectrometer covers the space between approximately 4.5 m and 11 m in radius and 7 m and 23 m longitudinally on both sides of the interaction point~\cite{ATLAS:1997ad}.
As a conservative estimation, we require $\eta$ to travel across the whole detector before decay to leave the tracks. The number of events with at least one track is calculated as
\begin{equation}\label{eq:Noftrk}
	N_{\rm trk}=\mathcal{L}\sigma(M_\eta)(2P_{\rm trk}-P_{\rm trk}^2),~
	P_{\rm trk}=
	\begin{cases}
		e^{-R /(d\sin\theta)} & |\tan\theta|>R/Z\\
		e^{-Z /(d\cos\theta)} & |\tan\theta|<R/Z
	\end{cases},
\end{equation}
where $d$ is the decay length of $\eta$, $R=11\, \mathrm{m} $ and $Z = 23\, \mathrm{m}$ are the radius and height of the outer boundary of the detector.

The $\eta$ production can be probed by existing $R$-hadron searches of the SUSY particle stop $\tilde t$, as they have the exact same quantum numbers.   Assuming that all the $\tilde t$ produced can fly across the whole muon spectrometer, the mass of $\tilde t$ is constrained to be larger than 1340 GeV with current $36.1\, {\rm fb}^{-1}$ LHC data~\cite{ATLAS:2019gqq}.
Matching Eq.~\eqref{eq:Noftrk} to the existing bound of stop $R$-hadron, we obtain the probed parameter space in $M_\eta$-$y$ plane as shown in Fig.~\ref{fig:dv_and_Rhadron} (right).
In our model, with increasing Yukawa coupling $y$, the lifetime of $\eta$ is shorter and fewer of them can travel across the detector before decay, resulting in a weaker lower bound on the mass of $\eta$.

\subsection{Asymmetry at muon collider}
The large CP asymmetry may lead to a smoking-gun signature at muon colliders that distinguishes our model experimentally from other scenarios without such large CP asymmetries. We construct the asymmetry observables at a high-energy muon collider, including the forward-backward asymmetry and charge asymmetry, respectively.

\subsubsection{Forward-backward asymmetry}

\begin{figure}
	\centering
	\includegraphics[width=0.52\textwidth]{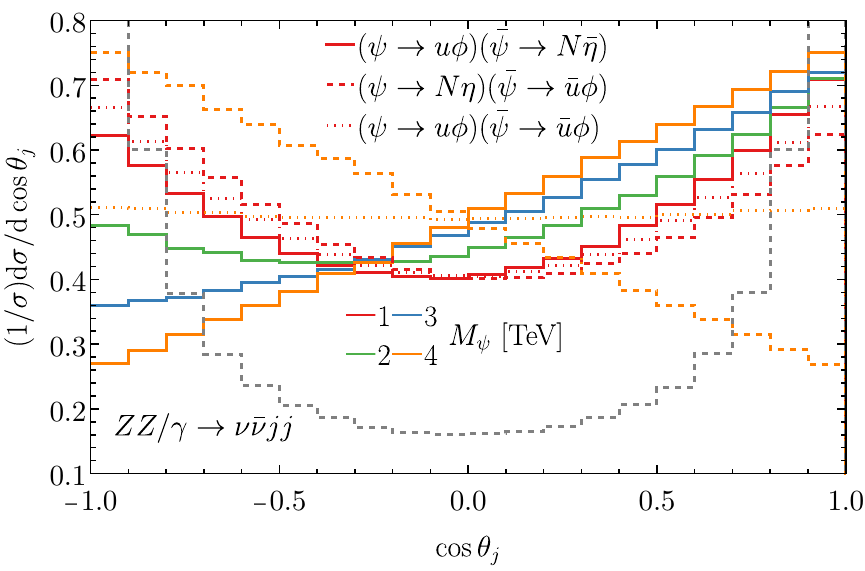}
	\includegraphics[width=0.47\textwidth]{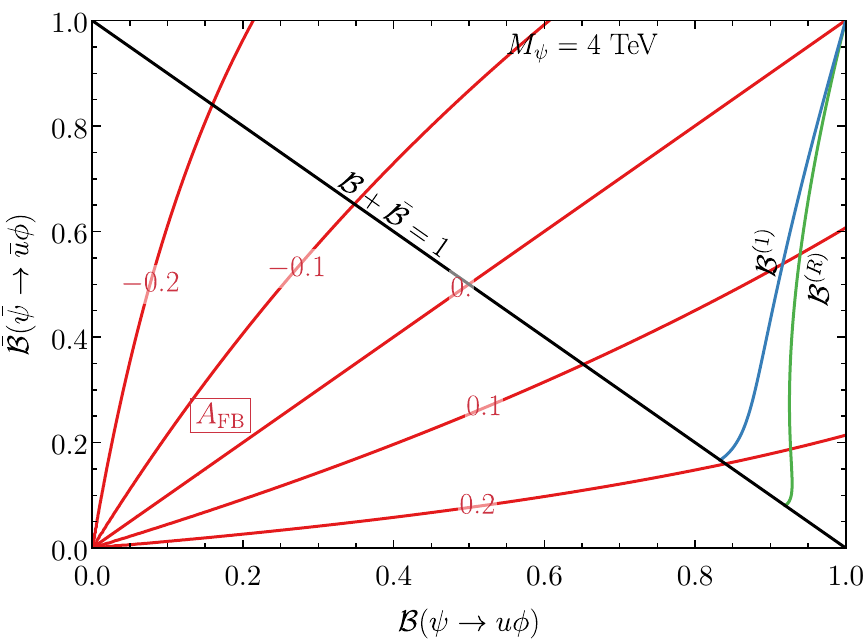}
	\includegraphics[width=0.49\textwidth]{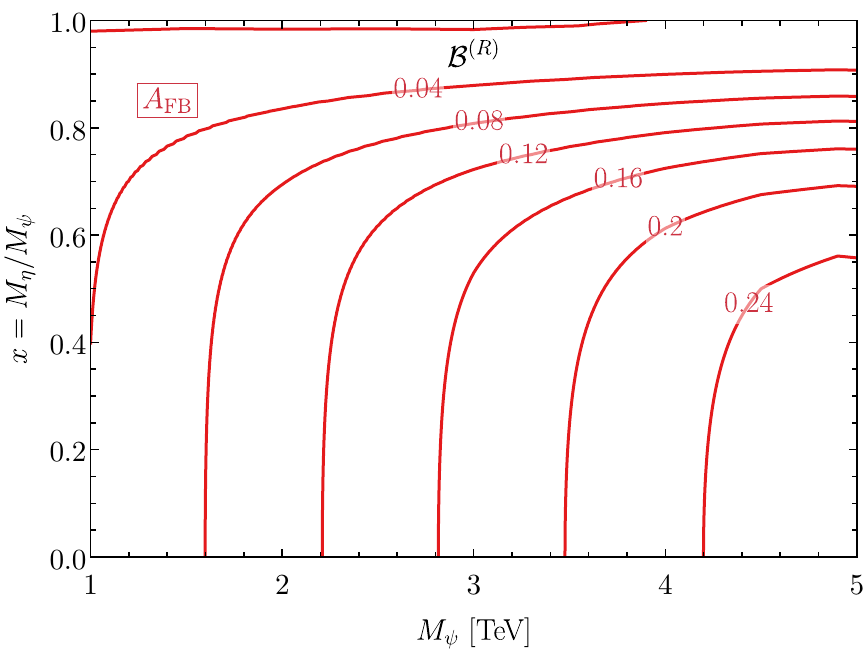}  
	\includegraphics[width=0.49\textwidth]{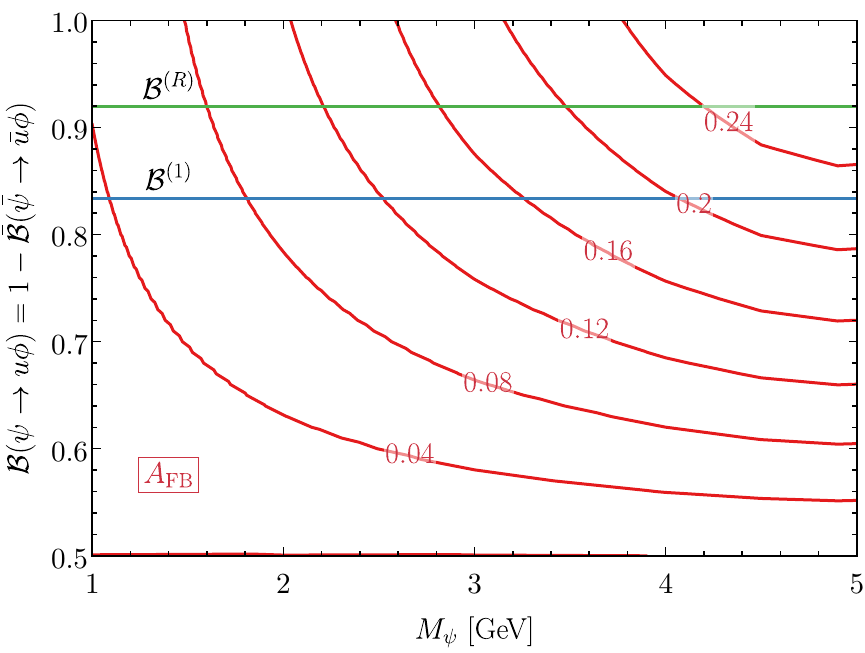} 
	\caption{For a 10-TeV muon collider:  Upper left: the leading jet angular distribution in the single and double jets of the $\psi\bar{\psi}$ pair production and decay. Upper right: the dependence of the forward-backward asymmetry $A_{\rm FB}$ on the branching fractions $\calB(\psi\to u\phi)$ and $\calB(\bar{\psi}\to\bar{u}\phi)$ with $M_\psi=4~\TeV$, and (lower left) on $M_\psi$ and $x=M_\eta/M_\psi$ with the resummed branching fractions $\calB^{(R)}$, (lower right) with the assumption that $\calB(\psi\to u\phi)=1-\calB(\bar{\psi}\to\bar{u}\phi)$.  Presented with both the first recursive branching fraction $\calB^{(1)}$ and resummed branching fractions $\calB^{(R)}$.
 }
	\label{fig:AFB}
\end{figure}

As indicated in Fig.~\ref{fig:Wpsi} in App.~\ref{app:loop}, the $\psi$ mainly decays into $u\phi$ and its anti-particle $\bar{\psi}$ mainly decays invisibly into $N\bar{\eta}$, which leaves the mono-jet signal. At a high-energy muon collider, the $\psi$ are produced through the pair production, either through the annihilation or vector boson (\emph{i.e.}, the hyper-charge gauge boson $B$) fusion as shown in Fig.~\ref{fig:feynMuC}, with cross-sections shown in Fig.~\ref{fig:xsLHC} (right). In the parameter space we are interested in, that is, $M_{\psi}\gtrsim1~\TeV$, the fusion process is suppressed due to the lower luminosity. We thus neglect their contribution. 
In lepton annihilation, it is well known that pair production via chiral couplings embraces a forward-backward asymmetry, $A_{\rm FB}$, defined as
\begin{equation}
A_{\rm FB}=\frac{\sigma_{+}-\sigma_{-}}{\sigma_{+}+\sigma_{-}},~
\sigma_{+}=\int_{0}^{1}\dd\cos\theta\frac{\dd\sigma}{\dd\cos\theta}, ~
\sigma_{-}=\int_{-1}^{0}\dd\cos\theta\frac{\dd\sigma}{\dd\cos\theta},
\label{eq:AFB}
\end{equation}
due to the different hyper-charges of left- and right-handed muons in the SM~\cite{ParticleDataGroup:2020ssz}. 
The difference between couplings of the left- and right-handed muons to the $Z$ boson leads to a non-zero spin component of the virtual $Z$ boson. This gives rise to the spin asymmetry of $\psi$, which propagates to the angular distribution of its decay products.
However, the specific value of $A_{\rm FB}$ depends on the $\psi$ mass, due to the washout in the boost kinematics.
As a consequence, we obtain that the $A_{\rm FB}$ of the mono-jet signal approximately linearly depends on $M_{\psi}$, as shown in the lower panel of Fig.~\ref{fig:AFB}.

In order to quantify the sensitivity of a high-energy muon collider on the $A_{\rm FB}$ in our model, we propose a measure defined as
\begin{equation}\label{eq:sensi}
\calS=\frac{N_+-N_{-}}{\sqrt{N_{\rm tot}+\epsilon^2 N_{\rm tot}^2}}, ~
N_{\rm tot}=N_{+}+N_{-}+N_{\rm SM},
\end{equation}
where $N_\pm=\calL\sigma_\pm$ are the event numbers in the forward and backward directions, respectively. 
Here we take the leading jet instead of the inclusive one, which will enhance the asymmetry due to the reduction of jet number in $N_++N_-$.
The $N_{\rm SM}$ is the SM background, which contributes to $A_{\rm FB}$ negligibly, to be shown later. The $\epsilon$ parameter is introduced to quantify the systematic uncertainty. Due to the clean environment at the muon collider as a lepton collider, we expect the systematic uncertainty can bring down to well under control. In some previous studies, a 0.1\% systematic uncertainty is taken as a nominal choice~\cite{Han:2020uak,Accettura:2023ked}. In this work, we vary $\epsilon$ between 0 and 1\% as optimistic and conservative scenarios, respectively.

\begin{figure}
    \centering
    \includegraphics[width=0.34\textwidth]{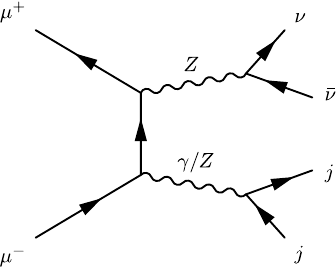}
    \caption{The fake background process to the mono-jet production at a high-energy muon collider.}
    \label{fig:feynBG}
\end{figure}

\begin{figure}
    \centering
\includegraphics[width=0.49\textwidth]{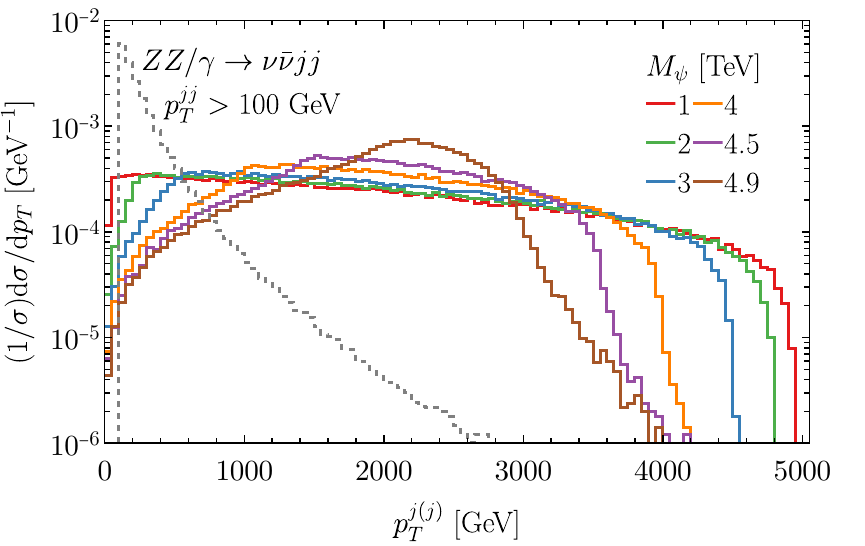}
    \includegraphics[width=0.48\textwidth]{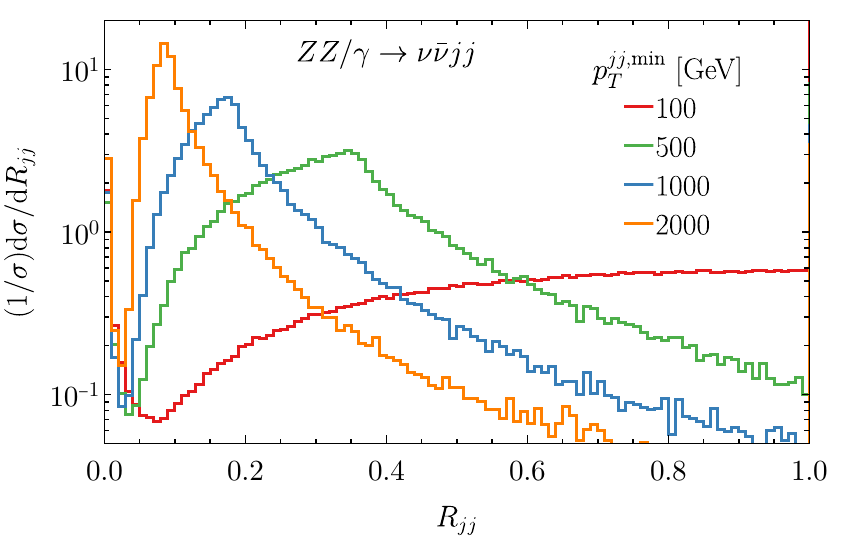} 
    \caption{For a 10-TeV muon collider: Left: the distribution of transverse momentum of the single jet and jet pair $p_T^{j(j)}$ for the signal $(\psi\to u\phi)(\bar{\psi}\to N\eta)$ and the faked background $ZZ/\gamma\to\nu\bar{\nu}jj$ (with a cut $p_T^{jj}>100~\GeV)$. Right: the separation distance between the two jets in the faked background with different minimal $p_T^{jj}$ cuts.}
    \label{fig:bkgdist}
\end{figure}

Naively thinking, the mono-jet signal at a high-energy muon collider is a background-free process. However, as the representative diagram in Fig.~\ref{fig:feynBG} shows, recoiled by a missing transverse energy in $Z\to\nu\bar{\nu}$, the di-jet production through a photon or $Z$ boson conversion can fake the background. 
In Fig.~\ref{fig:bkgdist} (left), we compare the distribution of transverse momentum of a jet pair $p_T^{jj}$ against the mono-jet signal $p_T^j$. On the right panel of Fig.~\ref{fig:bkgdist}, we show the di-jet separation distance $R_{jj}=\sqrt{(\Delta\phi_{jj})^2+(\Delta\eta_{jj})^2}$ in this fake background with different $p_T^{jj}$ cuts.
We see that when $p_T^{jj}>1~\TeV$, the di-jet separation peaks around $R_{jj}\sim2M_{Z}/p_T^{jj,\min}<0.2$, which would be very difficult to differentiate from the mono-jet signal.
In addition, we have a side peak for $R_{jj}\to 0$, which reflects the collinear $\gamma\to q\bar{q}$ splitting.

\begin{figure}
\centering
\includegraphics[width=0.49\textwidth]{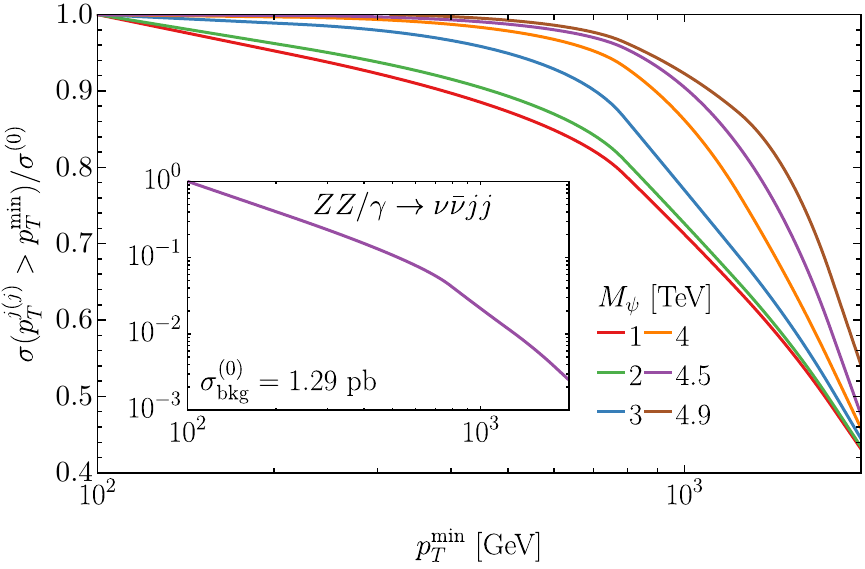}
\includegraphics[width=0.49\textwidth]{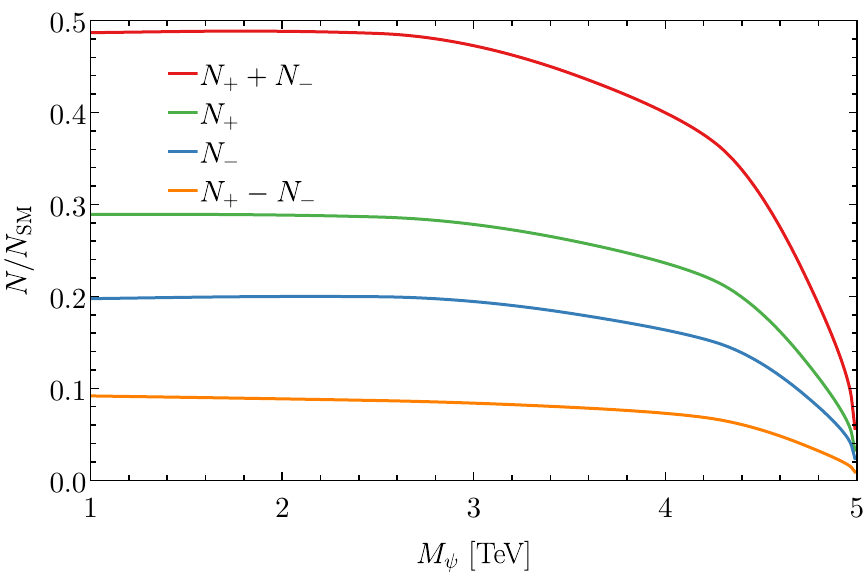}
\includegraphics[width=0.49\textwidth]{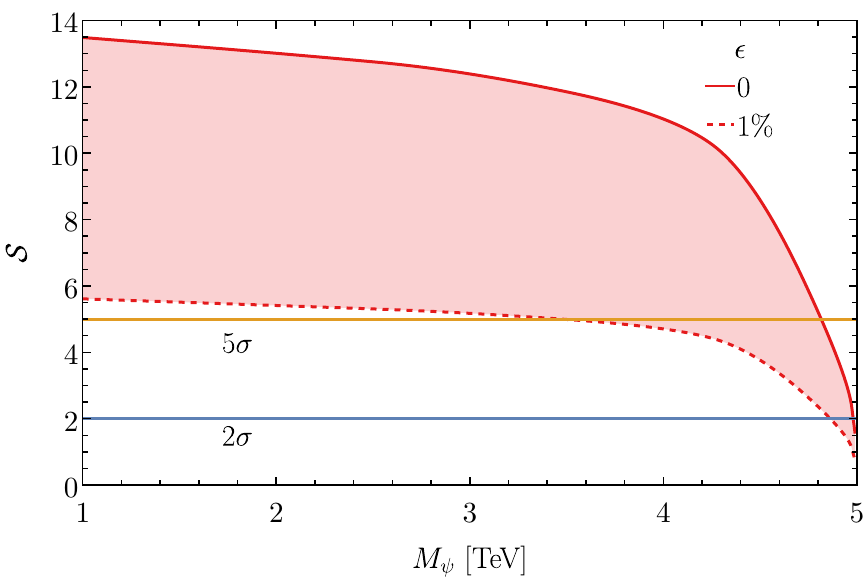}
\includegraphics[width=0.49\textwidth]{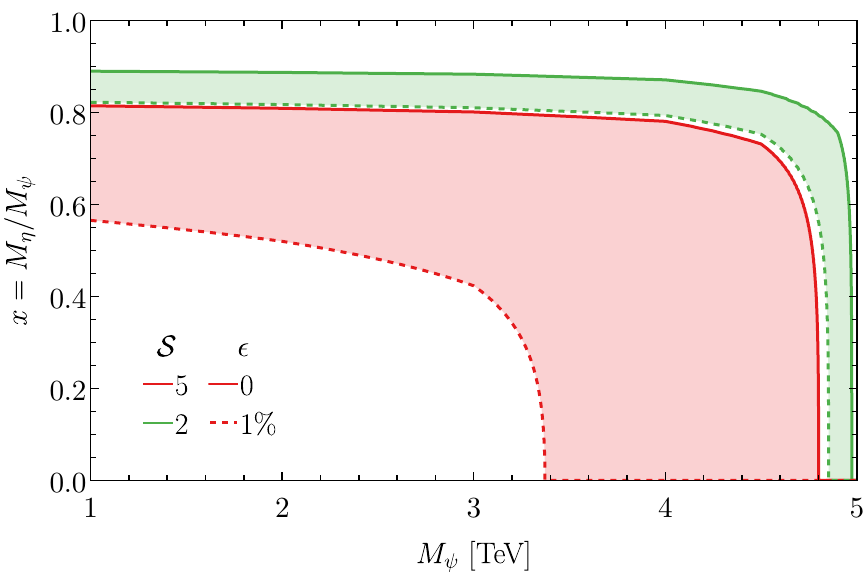}
\caption{For a 10-TeV muon collider:  Upper left: the efficiency of increasing minimal $p_T^{j(j)}$ cut for the signal and background, respectively. Upper right: the event numbers for the forward $N_+$ and backward $N_-$ together with their sum and difference, when normalized to the SM background with the optimization cut $p_T^{j(j)}>2~\TeV$. Lower: the measure of sensitivity $\calS$ for $M_\eta\ll M_\psi$ (left) and the $2(5)\sigma$ exclusion limit in the $(M_\psi,x=M_\eta/M_\psi)$ plane with an assumed $\epsilon=0(1\%)$ systematic uncertainty in terms of Eq.~(\ref{eq:sensi}).}
\label{fig:sensAFB}
\end{figure}

When examining the transverse distribution in Fig.~\ref{fig:bkgdist} (left) carefully, we see that the background dies out very rapidly with respect to signals. It gives us an opportunity to optimize the signal-background ratio by increasing the transverse momentum cut $p_T^{\min}$. In Fig.~\ref{fig:sensAFB} (upper left), we show the cut efficiency for both background and signals with a few representative values of $M_{\psi}$.
We see that by increasing the $p_T^{\min}$ cut up to 2 TeV, the background is reduced by three orders of magnitude, while the signal only gets a reduction by about a factor of two, both consistent with the $p_T$ distribution in Fig.~\ref{fig:bkgdist} (left).
With this optimization cut $p_T^{j(j)}>2~\TeV$, we show the signal forward $N_{+}$ and background $N_{-}$ event numbers as well as their sum and difference, normalized to the SM fake background in Fig.~\ref{fig:sensAFB} (upper right). 
As a reminder, the SM background process $ZZ/\gamma\to\nu\bar{\nu}jj$ gives an inclusive cross-section of about $\sigma=3.09~\textrm{pb}$. With the cut efficiency in the upper left panel of Fig.~\ref{fig:sensAFB}, we obtain the SM fake background events as $N_{\rm SM}=3.2\times10^{4}$ with an integrated luminosity $\mathcal{L}=10~\textrm{ab}^{-1}$~\cite{Bartosik:2020xwr,Accettura:2023ked}.
In comparison, the signal gives forward $N_+$ and backward $N_-$ events about a factor of 0.3 and 0.2 with respect to the SM background $N_{\rm SM}$, shown in Fig.~\ref{fig:sensAFB} (upper right). As a result, we obtain a sizeable event number for the difference $N_{+}-N_{-}$. When folding in the measure in Eq.~(\ref{eq:sensi}), we obtain the $A_{\rm FB}$ sensitivity greater than 5 units when $3.5<M_{\psi}<4.7~\TeV$ in the limit $M_\eta\ll M_\psi$ with a conservative 1\% or optimistic negligible systematic uncertainty, as shown in Fig.~\ref{fig:sensAFB} (lower left).
It means that we can either discover the $\psi$ in this model up to a $5\sigma$ level in this range, or exclude the parameter space $M_{\psi}<3.5\sim4.7~\TeV$ at a 10 TeV muon collider. 
For completeness, we also present the sensitivity dependence on the mass ratio $x=M_\eta/M_\psi$ in Fig.~\ref{fig:sensAFB} (lower right), as a consequence of the mass-dependent asymmetry and branching fraction of Fig.~\ref{fig:AuBR}. With the increase in the relative $\eta$ mass, the asymmetry gradually dies out, and the corresponding sensitivity to $\psi$ mass becomes weaker.

\subsubsection{Charge asymmetry}

As mentioned above, the $A_{\rm FB}$ is not unique for our model. Instead, the asymmetry in Eq.~(\ref{eq:Au}) ensures that $\psi$ almost decays visibly while its antiparticle $\bar{\psi}$ almost decays invisibly, which gives a distinctive feature.
As a consequence, the visible decay product (mainly the $u$ quark here) will carry the charge and color information of the mother particle $\psi$. Sequentially, the $u$-quark will shower and then hadronize into baryons and mesons, which will deposit energy into the hadronic calorimeter.
Here we use \texttt{Pythia8}~\cite{Bierlich:2022pfr} to simulate the parton shower and hadronization. The colored and charged scalar $\eta$ will fly out of the detectors, due to its long-lived feature, which only leaves colored tracks as explored in Sec.~\ref{sec:DV}. It will finally hadronize into $R$-hadrons, which is taken into account by \texttt{Pythia8}, but is not relevant here.

\begin{figure}
\includegraphics[width=0.53\textwidth]{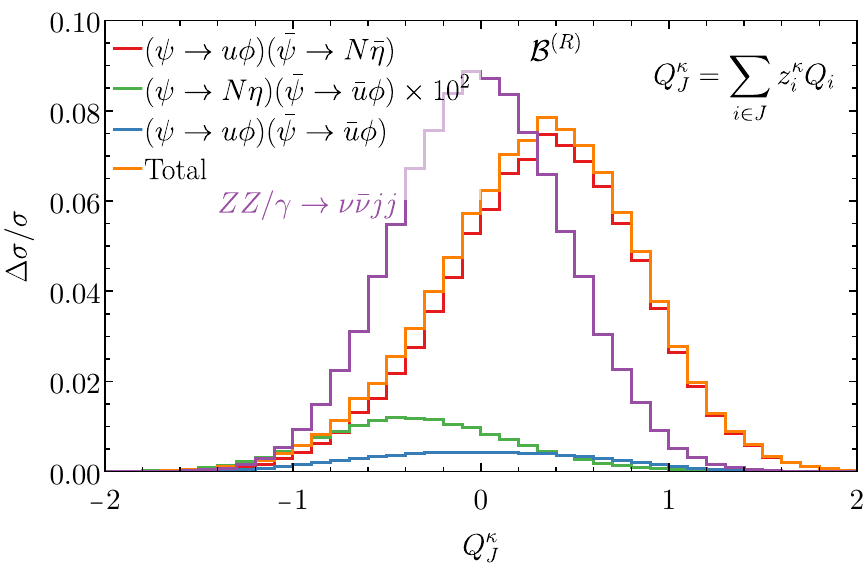}
\includegraphics[width=0.46\textwidth]{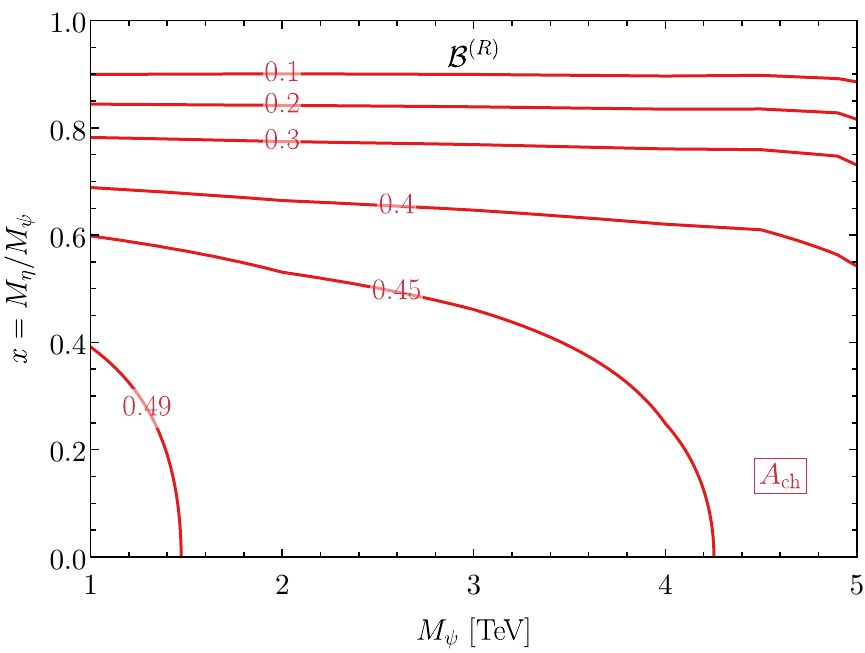}
\includegraphics[width=0.49\textwidth]{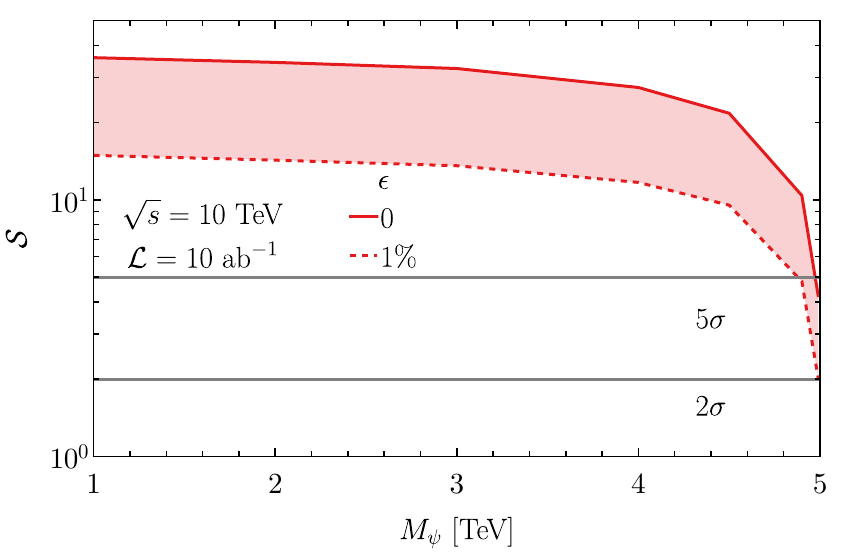}
\includegraphics[width=0.49\textwidth]{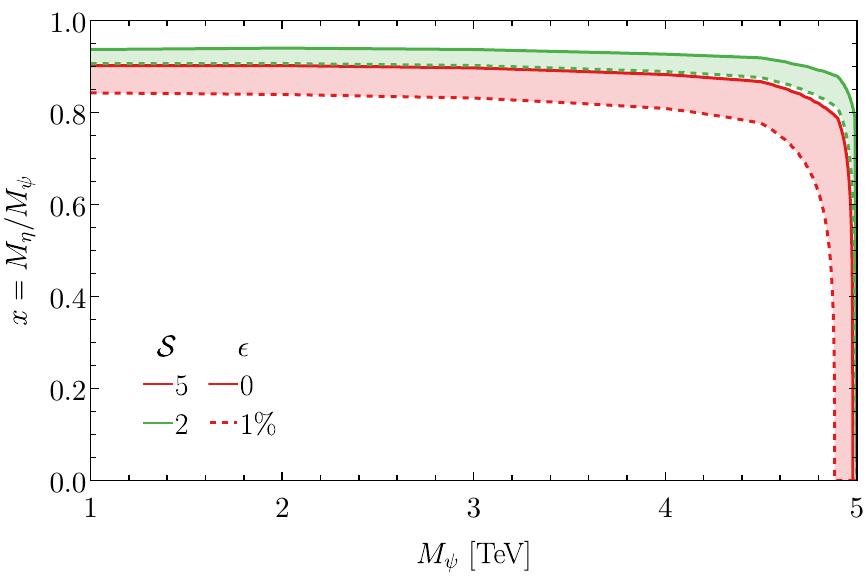}
\caption{For a 10-TeV muon collider: Upper left: the leading jet charge distribution in $\psi\bar{\psi}$ pair production. 
Upper right: the sensitivity based on the jet charge asymmetry. 
Lower: similar to Fig.~\ref{fig:sensAFB} (lower) but for the charge asymmetry.}
\label{fig:jetcharge}
\end{figure}

In Fig.~\ref{fig:jetcharge} (upper left), we show the jet charge of the signal mono-jets as well as the faked background $ZZ/\gamma\to\nu\bar{\nu}jj$. The jet charge is defined as~\cite{Field:1977fa,Krohn:2012fg}
\begin{equation}\label{eq:jetcharge}
Q_J^{\kappa}=\sum_{i\in J}z_i^\kappa Q_i, ~ z_i=p_T^i/p_T^{J},
\end{equation}
with the choice $\kappa=0.3$~\cite{Li:2021uww,Li:2023tcr,Kang:2023ptt}. Here, $p_T^i$ and $Q_i$ are the transverse momentum and electric charge of each jet constituent, and $p_T^J$ denotes the transverse momentum of the jet based on the anti-$k_T$ algorithm with the choice of $\Delta R=0.4$.
Also, instead of the inclusive jets, we only consider the charge of the leading $p_T$ jet here. In Fig.~\ref{fig:jetcharge} (upper left), we see that the SM fake background gives the peak of $Q_J^\kappa=0$, while the signals give a positive peak around $Q_{J}^{\kappa}\sim0.4$, consistent with the up-type jet in Ref.~\cite{Li:2021uww}. Also, we obtain a very little dependence on the $\psi$ mass, reflecting the charge feature as a quantum number instead of a kinematic variable. Based on the distribution in Fig.~\ref{fig:jetcharge} (upper left), we can construct a jet charge asymmetry $A_{\rm ch}$
with a similar definition as Eq.~(\ref{eq:AFB}), by only changing $\sigma_\pm(N_\pm)$ to the positive/negative jet charge cross-sections (event numbers), respectively, with the result shown in Fig.~\ref{fig:jetcharge} (upper right).
As expected, the $A_{\rm ch}$ has a very small dependence on the $\psi$ mass. With the increase of $M_\psi$, the $A_{\rm ch}$ decreases slightly, probably because a smaller velocity in the $\psi$ decay involves more hadronic activities from the vacuum, which retain less charge information of the mother particle.
Based on this asymmetry, we can project the sensitivity in terms of Eq.~(\ref{eq:sensi}) as well, with the result shown in Fig.~\ref{fig:jetcharge} (lower). We see that we have a very good discrimination of the $\psi$ up to $M_{\psi}\sim4.9~\TeV$ even with the 1\% systematic uncertainty at a 10 TeV muon collider. That is to say, we can either discover $\psi$ or exclude this model within $M_{\psi}<4.9~\TeV$ once the muon collider is turned on.

\section{Summary and Conclusion}
\label{sec:conclusion}
Understanding the baryon and antibaryon imbalance in the observed universe remains one of the most challenging puzzles in contemporary particle physics and cosmology~\cite{ParticleDataGroup:2020ssz,Planck:2018vyg}. 
In this work, we proposed a minimal baryogenesis scenario, dubbed Dirac baryogenesis, motivated by the Dirac leptogenesis~\cite{Dick:1999je,Murayama:2002je}. As summarized in Table \ref{tab:particle} of Section \ref{sec:model}, we introduce double copies of vector-like colored fermion $\psi_i\ (i=1,2)$, a colored charged scalar $\eta$, as well as a neutral fermion $N$ and scalar $\phi$, with the corresponding Lagrangian in Eq.~(\ref{eq:Lag}). With the decays of heavy $\psi$'s in Fig.~\ref{fig:AuBR} (right), equal and opposite CP asymmetries are created in two sectors, $u\phi$ and $N\eta$, respectively, which are prevented from equilibrating with each other at the early Universe. As explained in Sec.~\ref{sec:loop} and detailed in App.~\ref{app:Asym}, an order of a unit $\mathcal{O}(1)$ CP asymmetry can be achieved with the resonant enhancement in Eq.~(\ref{eq:dMpsi}) together with the maximal Yukawa phase in Eq.~(\ref{eq:Y}). 

With benchmark points in Table \ref{tab:benchmark} of Section \ref{sec:cosmo}, the evolution of the comoving abundance $Y_i=n_i/s$ of baryon asymmetry is demonstrated in Fig.~\ref{fig:BG} (left). We see that the strong washout and dilution of the asymmetry can be balanced with a suitable choice of parameters, and thus successful baryogenesis can be achieved at TeV scale. By fixing the mass ratio $x=M_\eta/M_\psi=0.1/0.5/0.8$, the desired parameter space $(M_\psi,y)$ consistent with observed baryon asymmetry $\Omega_bh^2= 0.0224\pm0.0005$ within $5\sigma$~\cite{Planck:2018vyg} is shown in Fig.~\ref{fig:BG} (right), which keeps the extended SM sector at the TeV scale. The unbroken $Z_4$ symmetry, considered to prevent unwanted terms, also allows a stable scalar singlet dark-matter candidate $\phi$, whose relic can be obtained primarily via Higgs portal annihilation.

Because of the need for the new particles around the TeV scale, we proposed to search for those particles, as well as their characteristic signature of large CP asymmetry at colliders, including the current LHC and future lepton colliders, in Sec.~\ref{sec:collider}.

Due to gauge and Yukawa interactions, the colored fermion $\psi$ can be produced through a pair or in association with the neutral scalar $\phi$, as shown in Fig.~\ref{fig:feynLHC}. As the particle $\psi$ largely decays into a visible up quark and $\phi$ while its antiparticle $\bar{\psi}$ mainly decays into invisible $N$ and $\eta$, both the pair and associated production of $\psi$ can give a mono-jet signal with a large missing transverse energy. With the existing measurement of mono-jet at the 13 TeV LHC~\cite{ATLAS:2017bfj}, we can exclude the heavy $\psi$ up to $M_\psi\lesssim1.5~\TeV$ in the absence of Yukawa coupling, $y_L=0$, shown in Fig.~\ref{fig:Prompt-Constrain} of Sec.~\ref{sec:monojet}. The single production only becomes dominant when the Yukawa coupling is large, \emph{e.g.}, $y_L=1$, and the exclusion limit can be extended to $M_{\psi}\gtrsim 2.4~\TeV$ and 3.5 TeV for the current LHC~\cite{ATLAS:2017bfj} and future HL-LHC measurements, respectively.

As the scalar $\eta$ also carries color charge, it can be easily produced in pairs through the QCD interactions. Considering its long-lived feature, as shown in Fig.~\ref{fig:decay}, it will leave the displaced vertices and colored tracks, as explored in Sec.~\ref{sec:DV}. 
As demonstrated with $y\sim10^{-3}$ and the virtual $\psi$ mass $M_{\psi}=5~\TeV$ in Fig.~\ref{fig:dv_and_Rhadron} (left), we can exclude $\eta$ up to $M_\eta\lesssim1.5~\TeV$, which can be extended to $M_\eta\lesssim2.2~\TeV$ with the accumulation of luminosity of 3000 fb$^{-1}$ at HL-LHC. Similar to other stable heavy particles, $\eta$ can also leave a colored track signature, which can exclude $\eta$ up to a similar mass region, but with a much smaller Yukawa coupling, as shown in Fig.~\ref{fig:dv_and_Rhadron}~(right).

Considering the feature of the CP asymmetry in baryogenesis, we also study the prospects of discovering these particles via observation of forward-backward and charge asymmetry at a future 10 TeV muon collider.
Due to the chiral coupling of the hyper-charge gauge boson $B$ with muon beams, the $\psi$-pair production through annihilation gives a natural spin asymmetry along the beam direction.
Considering different dominant decay channels of the $\psi$ particle and its antiparticle, the spin asymmetry will be inherited by its decay product, \emph{i.e.}, the visible mono-jet, which can be directly measured at muon colliders.
With our proposed asymmetry measure in Eq.~(\ref{eq:sensi}), we optimize the jet transverse momentum cut, and obtain an exclusion limit of the $\psi$ mass up to $M_\psi\lesssim4.8\ (3.4)~\TeV$ with $M_\eta\ll M_\psi$ and an assumed 0\ (1\%) systematic uncertainty, as shown in Fig.~\ref{fig:sensAFB}. Similarly, with the jet charge observable $Q_J^\kappa$ in Eq.~(\ref{eq:jetcharge}), we can directly measure the charge asymmetry $A_{\rm ch}$ in terms of the final-state mono-jet. With a similar measure, we can exclude the $\psi$ mass up to $M_\psi\lesssim4.9~\TeV$, very close to the collider threshold reach $\sqrt{s}/2$.

\begin{figure}
\centering
\includegraphics[width=0.47\textwidth]{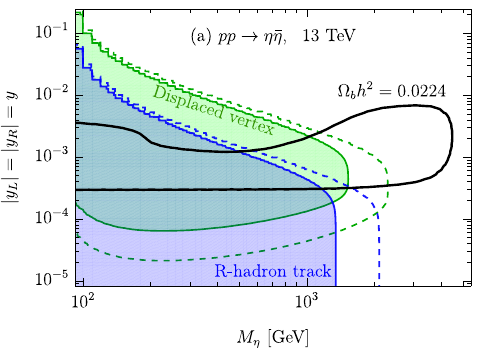}   
\includegraphics[width=0.52\textwidth]{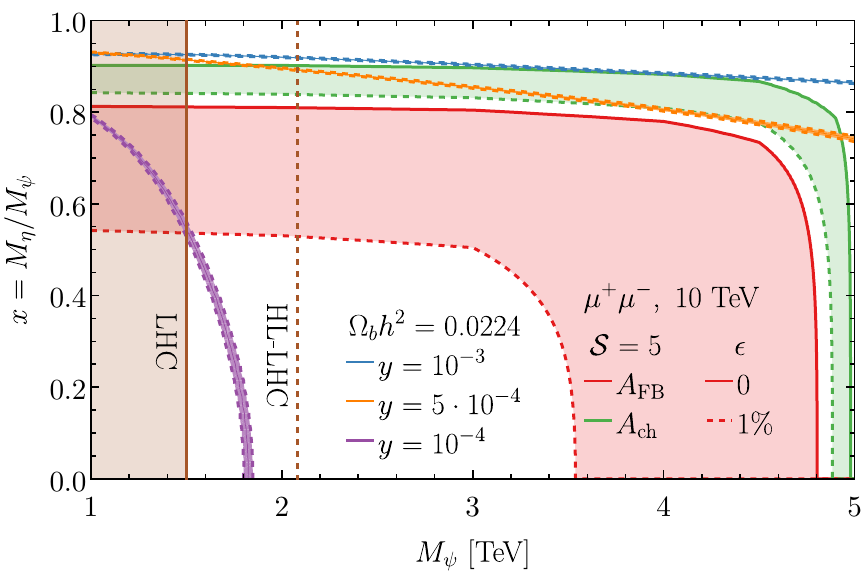}
    \caption{Summary plots: Left: The parameter $(M_\eta,y)$ that gives correct baryon asymmetry with $M_\psi=5~\TeV$ (black line), together with the parameter space that has been ruled out from displaced vertex (green shaded region) and $R$-hadron track (blue shaded region) at current LHC, and future HL-LHC projections with a luminosity of $3000~\text{fb}^{-1}$ (dashed contour). 
    Right: The parameter space $(M_\psi,x=M_\eta/M_\psi)$ probed by a 10 TeV muon collider with the forward-backward asymmetry ($A_{\rm FB}$) and charge asymmetry ($A_{\rm ch}$) in comparison with the baryon asymmetry $\Omega_bh^2=0.0224\pm0.0005$ within 5$\sigma$~\cite{Planck:2018vyg}. The solid vertical line is ruled out by the LHC mono-jet search~\cite{ATLAS:2017bfj}, while the dashed line denotes the HL-LHC projection.}
    \label{fig:money}
\end{figure}

In Fig.~\ref{fig:money}, we summarize the collider exclusion limits from the LHC displaced vertices, colored tracks, as well as forward-backward and charge asymmetry at a 10 TeV muon collider together with the constraints from observed baryon asymmetry, demonstrating the complementarity between the collider and cosmological probes. As the left panel of Fig.~\ref{fig:money} shows, a major part of the $(M_\eta, y)$ parameter space consistent with successful baryogenesis (solid black line) is already being ruled out by the LHC while HL-LHC can probe some more parts of it via displaced vertex and $R$-hadron track signatures of long-lived $\eta$. Similarly, the (HL-)LHC's the mono-jet with missing energy measurement and a future muon collider can probe the parameter space consistent with baryogenesis via forward-backward and jet charge asymmetries with $\geq 5\sigma$ sensitivity, as shown on the right panel of Fig. \ref{fig:money}. In the limit of resonantly enhanced CP asymmetry, it is indeed possible to distinguish decay asymmetries into particles and anti-particles at colliders, providing a direct probe of a low-scale baryogenesis mechanism. 

While we focus on a minimal baryogenesis mechanism in this work and its smoking-gun signature of decay asymmetries, this novel collider probe can be utilised for any low-scale baryogenesis or leptogenesis mechanisms in the resonant limit with $\mathcal{O}(1)$ CP asymmetry. Given that generic low-scale baryogenesis scenarios require resonantly enhanced CP asymmetry, our study provides a unique way of probing the origin of matter-antimatter asymmetry directly at colliders.

\begin{acknowledgements}
We would like to thank Bin Yan for providing the jet charge example code and a helpful discussion.
This work is supported by the U.S. Department of Energy under grant No. DE-SC0007914, and also in part by the Pitt PACC. K.X. is also supported by the U.S. National Science Foundation under Grants No. PHY-2112829, PHY-2013791, and PHY-2310497. 
The work of T.H. and K.X. was performed partly at the Aspen Center for Physics,
which is supported by the U.S. National Science Foundation under Grant No. PHY-1607611
and No. PHY-2210452. 
The work of D.B. is supported by the Science and Engineering Research Board (SERB), Government of India grants MTR/2022/000575 and the Fulbright-Nehru Academic and Professional Excellence Award 2024-25. 
\end{acknowledgements}

\appendix

\section{Loop induced asymmetry}
\label{app:Asym}

In this section, we lay out the details of the calculations related to the generation of bayron asymmetry, through one-loop decay of $\psi$. Meanwhile, we will also discuss the condition of the asymmetry maximization, which is mainly adopted in this work.

\subsection{Loop induced asymmetry}
\begin{figure}
\includegraphics[width=0.3\textwidth]{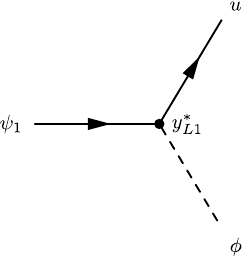}
\includegraphics[width=0.4\textwidth]{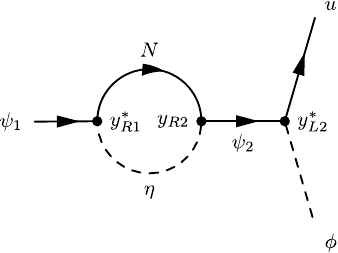}
\caption{Dominant processes sourcing the CP asymmetry in SM quark sector.}
\label{fig:LEPTO}
\end{figure}

As constructed in Sec.~\ref{sec:model}, the baryon number asymmetry is generated through the out-of-equilibrium decay of $\psi\to u\phi$ after taking radiative corrections into account, as depicted in Fig.~\ref{fig:LEPTO}. We can define an asymmetry as
\begin{equation}\label{eq:delta}
\delta_u=\frac{\Gamma(\psi\to u\phi)-\Gamma(\bar{\psi}\to\bar{u}\bar{\phi})}{\Gamma(\psi\to u\phi)+\Gamma(\bar{\psi}\to\bar{u}\bar{\phi})}.
\end{equation}
At the one-loop level, the decay width can be expressed in terms of
\begin{equation}
\Gamma(\psi\to u\phi)=\Gamma^{(0)}|1+R|^2, ~
\Gamma(\bar{\psi}\to \bar{u}\bar{\phi})=\Gamma^{(0)}|1+\bar{R}|^2,
\end{equation}
where $\Gamma^{(0)}$ denotes the tree-level decay width, identical for $\psi\to u\phi$ and $\bar{\psi}\to\bar{u}\phi$. $R$ and $\bar{R}$ are the loop corrections to $\psi$ and $\bar{\psi}$ decay amplitudes in Fig.~\ref{fig:LEPTO}, which can be expressed as
\begin{equation}
R=R_cR_{\calM}, ~\bar{R}=R_c^*R_{\calM},
\end{equation}
with $R_c$ denoting coupling coefficients and $R_{\calM}$ is the loop amplitude factors. Pay attention that $\psi$ and $\bar{\psi}$ share the same loop amplitude due to $\bar{u}u=\slashed{p}=\bar{v}v$~\cite{Davidson:2008bu}.
As a result, we can express the asymmetry as
\begin{equation}\label{eq:R4delta}
\delta_u=\frac{R_{\calM}(R_c-R_c^*)+\hc}{2(1+|R|^2)+R_\calM(R_c+R_c^*)+\hc}.
\end{equation}
Just a side remark that, there exists no triangle or one-loop vertex corrections that contribute to the CP asymmetry in our model at leading order, and hence they are not considered here.

The one-loop amplitude ratio $R$ for $\psi_1\to u\phi$ decay can be constructed in terms of the $N\eta$ and $u\phi$ loops respectively as
\begin{equation}\label{eq:R}
\begin{aligned}
&R_{\psi_1\to u\phi}^{N\eta}=\frac{\mathcal{M}^{(1)}(\psi_1\xrightarrow{N\eta} u\phi)}{\mathcal{M}^{(0)}(\psi_1\to u\phi)}
=\frac{i\Im B_1(M_{\psi_1}^2,M_N^2,M_\eta^2)}{16\pi^2y_{L1}}
\sum_{j\neq1}\frac{M_{\psi_1}M_{\psi_j}y_{R1}^*y_{Rj}y_{Lj}}{M_{\psi_1}^2-M_{\psi_j}^2+iM_{\psi_j}\Gamma_{\psi_j}},\\
&R_{\psi_1\to u\phi}^{u\phi}=\frac{\mathcal{M}^{(1)}(\psi_1\xrightarrow{u\phi} u\phi)}{\mathcal{M}^{(0)}(\psi_1\to u\phi)}
=\frac{i\Im B_{1}(M_{\psi_1}^2,M_u^2,M_\phi^2)}{16\pi^2}
\sum_{j\neq1}\frac{M_{\psi_1}^2y_{Lj}^*y_{Lj}}{M_{\psi_1}^2-M_{\psi_j}^2+iM_{\psi_j}\Gamma_{\psi_j}}.
\end{aligned}
\end{equation}
Here, $B_1$ is the Passarino-Veltman loop function defined as
\begin{equation}
\slashed{p} B_1\left(p^2, m_A^2, m_B^2\right)=\int \frac{d^4 k}{i \pi^2} \frac{-\slashed{k}}{\left(k^2-m_A^2\right)\left((p-k)^2-m_B^2\right)}.
\end{equation}
The $j=1$ loop diagram corresponds to the $\psi_1$ self energy, which is absorbed into the wave-function/mass renormalization~\cite{Pilaftsis:1997jf,Pilaftsis:1998pd,Pilaftsis:2003gt}.
We see that the $u\phi$ loop does not contribute to the asymmetry $\delta$ due to the zero difference between $R_c=y_{Lj}y_{Lj}^*$ and its conjugate $R_c^*$.

The asymmetry mainly comes through the $N\eta$ loop as 
\begin{equation}
\begin{aligned}
\delta_u
&\sim R_{\calM}\frac{R_c-R_c^*}{2}+\hc\\
&=\sum_{j\neq1}\frac{-i\Im B_1(M_{\psi_1}^2,M_N^2,M_\eta^2)M_{\psi_1}M_{\psi_j}}{16\pi^2(M_{\psi_1}^2-M_{\psi_j}^2+iM_{\psi_j}\Gamma_{\psi_j})}
\frac{y_{R1}^*y_{Rj}y_{Lj}y_{L1}^*-y_{R1}y_{Rj}^*y_{Lj}^*y_{L1}}{2y_{L1}y_{L1}^*}+\hc.
\end{aligned}
\end{equation}
In terms of the optical theorem, we can relate the $N\eta$ loop factor to the decay width as
\begin{equation}
-\Im B_1(M_{\psi_1}^2,M_N^2,M_\eta^2)=\frac{16\pi^2\Gamma(\psi_1\to N\eta)}{|y_{R1}|^2M_{\psi1}}.
\end{equation}
Then, the asymmetry can be expressed as
\begin{equation}
\delta_u\sim2\frac{\Im[y_{R1}^*y_{Rj}y_{Lj}y_{L1}^*]}{|y_{R1}|^2|y_{L1}|^2}\frac{\Gamma(\psi_1\to N\eta)M_{\psi_j}(M_{\psi_1}^2-M_{\psi_j}^2)
}{(M_{\psi_1}^2-M_{\psi_j}^2)^2+M_{\psi_j}^2\Gamma_{\psi_j}^2},
\end{equation}
which shares a similar form in Ref.~\cite{Pilaftsis:2003gt}.

We can maximize the asymmetry $\delta$ by taking
\begin{equation}\label{eq:dMpsi}
\frac{\Delta M_{\psi}^2}{M_\psi^2}\sim\frac{2\Delta M_\psi}{M_\psi}=\frac{\Gamma_{\psi}}{M_\psi}\implies
\delta_{\max}\sim\frac{\Im[y_{R1}^*y_{Rj}y_{Lj}y_{L1}^*]}{|y_{R1}|^2|y_{L1}|^2}\calB^{(0)}(\psi_1\to N\eta),
\end{equation}
corresponding the results in Refs.~\cite{Pilaftsis:1997jf,Pilaftsis:1998pd}.
Meanwhile, we can choose 
\begin{equation}\label{eq:Y}
y_{L1}=y_{L2}=iy_{R1}=y_{R2}=y
\end{equation}
to maximize the CP violating phase. This Yukawa choice also gives a zero for $R_c+R_c^*$ in the denominator in Eq.~(\ref{eq:R4delta}).
Under such a condition, the resulting asymmetry at the zero-th order behaves as the branching fractions in Fig.~\ref{fig:decay} (left).

As a final remark in this subsection, the parameter choice in Eq.~(\ref{eq:Y}) will give a same asymmetry for the $\psi_2$. In addition, we will also obtain an opposite asymmetry in the $\psi\to N\eta$ and $\bar{\psi}\to N\bar{\eta}$ decay channel as
\begin{equation}\label{eq:deta}
\delta_\eta=\frac{\Gamma(\psi_1\to N\eta)-\Gamma(\bar{\psi}_1\to N\bar{\eta})}{\Gamma(\psi_1\to N\eta)+\Gamma(\bar{\psi}_1\to N\bar{\eta})}\sim
-\calB^{(0)}(\psi_1\to u\phi).
\end{equation}

\subsection{The one-loop decay widths and branching fractions}
\label{app:loop}
With the mass difference in Eq.~(\ref{eq:dMpsi}) as well as the Yukawa choice in Eq.~(\ref{eq:Y}), we have the loop ratios as
\begin{equation}
\begin{aligned}
&R^{u\phi}_{\psi\to u\phi}=\frac{\Gamma^{(0)}(\psi_1\to u\phi)}{\Gamma_{\psi_2}}\frac{-1+i}{2},~&
&R^{N\eta}_{\psi\to u\phi}=\frac{\Gamma^{(0)}(\psi_1\to N\eta)}{\Gamma_{\psi_2}}\frac{1+i}{2},\\
&R^{u\phi}_{\psi\to N\eta}=\frac{\Gamma^{(0)}(\psi_1\to u\phi)}{\Gamma_{\psi_2}}\frac{-1-i}{2},~&
&R^{N\eta}_{\psi\to N\eta}=\frac{\Gamma^{(0)}(\psi_1\to N\eta)}{\Gamma_{\psi_2}}\frac{-1+i}{2},
\end{aligned}
\end{equation}
where the width $\Gamma_{\psi_2}$ is with respect to the complex mass of $\psi_2$ in the propagator, which can include all-order resummation and differ from the leading order total width defined as
\begin{equation}\label{eq:GammaLO}
\Gamma_{\psi_2}^{(0)}=\Gamma^{(0)}(\psi_2\to u\phi)+\Gamma^{(0)}(\psi_2\to N\eta).
\end{equation}
Summing over both $u\phi$ and $N\eta$ loops gives the total one-loop correction to $\psi_1\to u\phi$ decay amplitude as
\begin{equation}
\begin{aligned}\label{eq:Rpsi}
&R_{\psi\to u\phi}=R^{u\phi}_{\psi\to u\phi}+R^{N\eta}_{\psi\to u\phi}
=\frac{-\Delta\Gamma_{\psi_1}^{(0)}+i\Gamma_{\psi_1}^{(0)}}{2\Gamma_{\psi_2}},\\
&R_{\psi\to N\eta}=R^{u\phi}_{\psi\to N\eta}+R^{N\eta}_{\psi\to N\eta}
=\frac{-\Gamma_{\psi_1}^{(0)}-i\Delta\Gamma_{\psi_1}^{(0)}}{2\Gamma_{\psi_2}}.
\end{aligned}
\end{equation}
Here, we define the leading order difference as
\begin{equation}
\Delta\Gamma_{\psi_1}^{(0)}=\Gamma^{(0)}(\psi_1\to u\phi)-\Gamma^{(0)}(\psi_1\to N\eta).
\end{equation}

\begin{figure}
    \centering
    \includegraphics[width=0.49\textwidth]{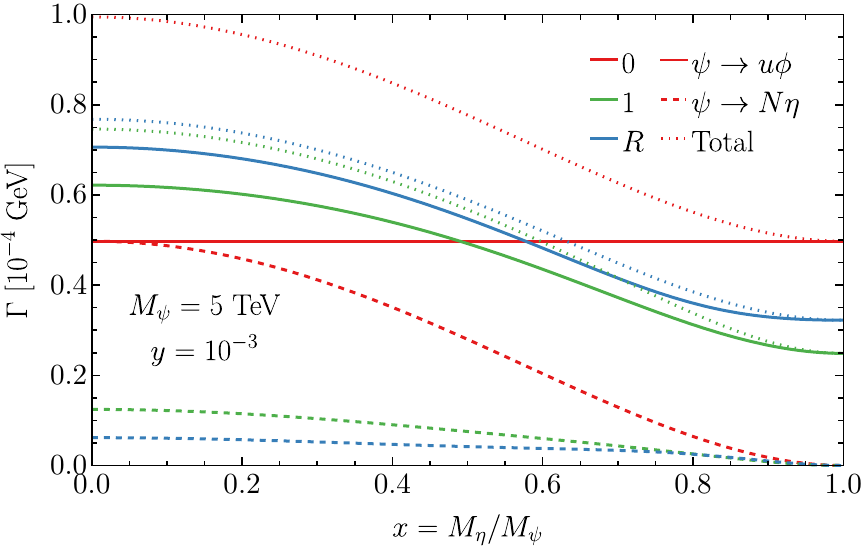}
    \includegraphics[width=0.49\textwidth]{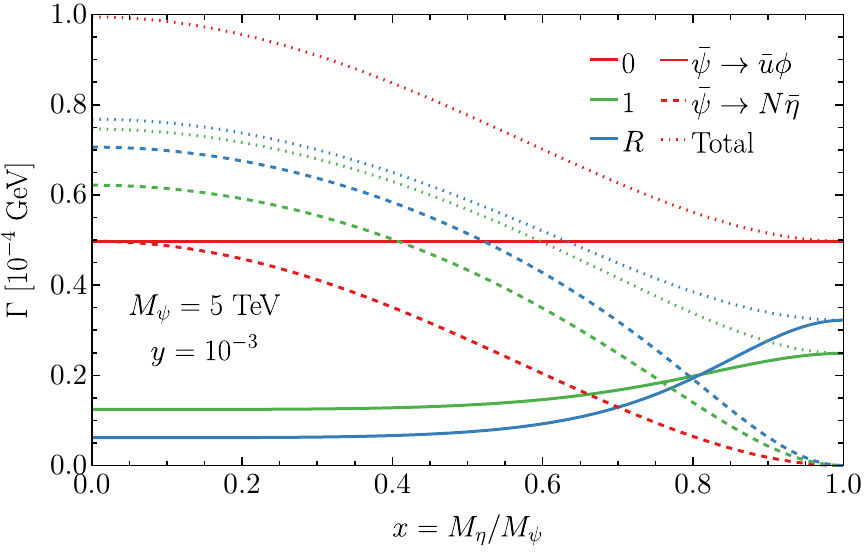}    
    \includegraphics[width=0.49\textwidth]{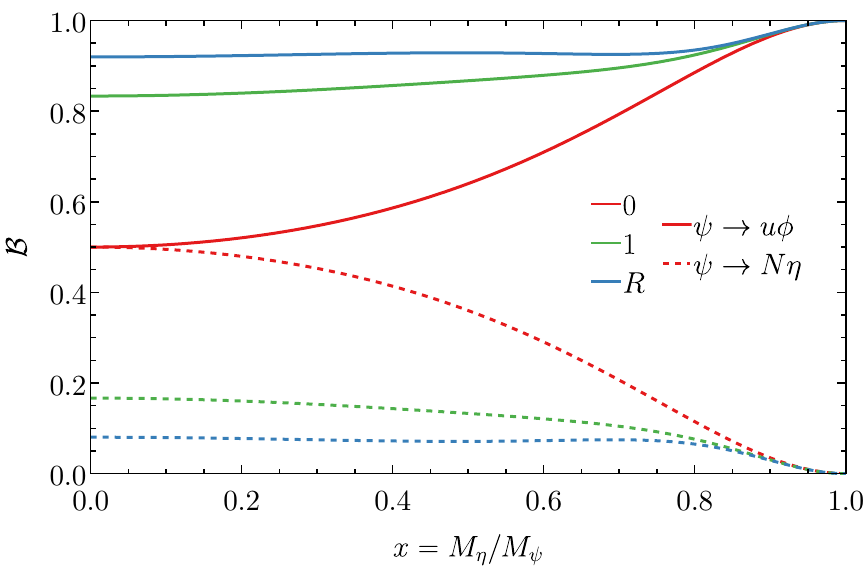}
    \includegraphics[width=0.49\textwidth]{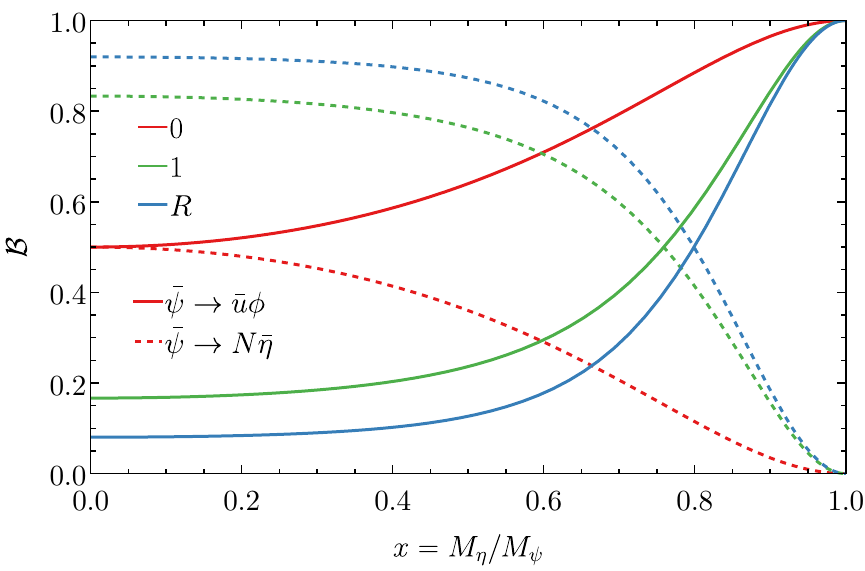}      
    \caption{The leading-order (0), first recursive (1) and resummed (R) decay widths (upper) and branching fractions (lower) for $\psi$ (left) and $\bar{\psi}$ (right), as a function of the mass ratio $x=M_\eta/M_\psi$.}
    \label{fig:Wpsi}
\end{figure}

\begin{figure}
    \centering
    \includegraphics[width=0.49\textwidth]{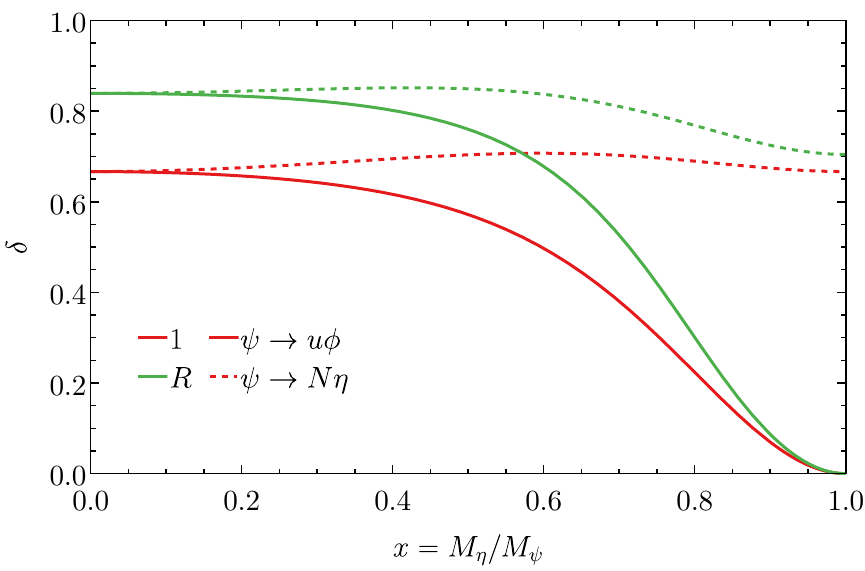}
    \caption{The first recursive (1) and resummed (R) asymmetry for $\delta_u$ and $-\delta_\eta$, as a function of the mass ratio $x=M_\eta/M_\psi$.}
    \label{fig:Asymmetry}
\end{figure}

The one-loop corrected partial width can be expressed as
\begin{equation}
\begin{aligned}
&\Gamma(\psi\to u\phi)=\Gamma^{(0)}(\psi\to u\phi)|1+R_{\psi\to u\phi}|^2,\\
&\Gamma(\psi\to N\eta)=\Gamma^{(0)}(\psi\to N\eta)|1+R_{\psi\to N\eta}|^2.
\end{aligned}
\end{equation}
The complete solution to the one-loop corrections can be achieved recursively starting with $\Gamma_{\psi_2}=\Gamma^{(0)}_{\psi_2}$ as Eq.~(\ref{eq:GammaLO}). 
The first recursion for the correction to $\psi\to N\eta$ partial width can be obtained with
\begin{equation}
|1+R_{\psi\to N\eta}|^2=1+2\Re[R_{\psi\to N\eta}]+|R_{\psi\to N\eta}|^2
=1-\frac{\Gamma_{\psi_1}^{(0)}}{\Gamma_{\psi_2}}+|R_{\psi\to N\eta}|^2,
\end{equation}
which gives $|R_{\psi\to N\eta}|^2$. To ensure this positivity, we will keep the loop squared term $|R|^2$ throughout this work, such as Eq.~(\ref{eq:R4delta}). We compare the first recursion for the decay widths and branching fractions with the leading-order ones in Fig.~\ref{fig:Wpsi}.
When $M_{\eta}\ll M_{\psi}$, we have 
\begin{equation}\label{eq:Gamma1}
\Gamma^{(1)}(\psi\to u\phi)=\frac{5}{4}\Gamma^{(0)},~
\Gamma^{(1)}(\psi\to N\eta)=\frac{1}{4}\Gamma^{(0)},~
\Gamma^{(1)}_{\psi}=\frac{3}{2}~\Gamma^{(0)}.
\end{equation}
Similarly, the one-loop corrected decay widths for the anti-particle when $M_\eta\ll M_\psi$ behaves as
\begin{equation}
\begin{aligned}
\Gamma^{(1)}(\bar{\psi}\to \bar{u}\bar{\phi})=\frac{1}{4}\Gamma^{(0)},~
\Gamma^{(1)}(\bar{\psi}\to N\bar{\eta})=\frac{5}{4}\Gamma^{(0)},~
\Gamma^{(1)}_{\bar{\psi}}=\frac{3}{2}~\Gamma^{(0)},
\end{aligned}
\end{equation}
which gives the corresponding asymmetry as
\begin{equation}
\delta_u=-\delta_\eta=\frac{2}{3}.
\end{equation}
The general dependence of $\delta_u$ and $\delta_\eta$ on the mass ratio $x=M_\eta/M_\psi$ can be found in Fig.~\ref{fig:Asymmetry}.

Beyond the recursion, the resummation for one-loop corrections can be achieved self-consistently by solving the equation
\begin{equation}
\Gamma_{\psi_2}=\Gamma_{\psi_1}=\Gamma(\psi_1\to u\phi)+\Gamma(\psi_1\to N\eta),
\end{equation}
as $\Gamma_{\psi_2}$ shows up on the right side in the denominator of Eq.~(\ref{eq:Rpsi}).
When $M_\eta\ll M_\psi$, we have the solution as
\begin{equation}
\Gamma_{\psi}^{(R)}=\frac{\Gamma_0}{3}\left[2-\frac{2}{(17+3\sqrt{33})^{1/3}}+(17+3\sqrt{33})^{1/3}\right]\simeq1.544\Gamma^{(0)},
\end{equation}
which is very close to the first recursive result in Eq.~(\ref{eq:Gamma1}), implying the fast convergence of the recursion. 
Meanwhile, we also have the corresponding partial widths as
\begin{equation}
\begin{aligned}
&\Gamma^{(R)}(\psi\to u\phi)=\Gamma^{(R)}(\bar{\psi}\to N\bar{\eta})=1.420\Gamma^{(0)},\\
&\Gamma^{(R)}(\psi\to N\eta)=\Gamma^{(R)}(\bar{\psi}\to \bar{u}\bar{\phi})=0.124\Gamma^{(0)},    
\end{aligned}
\end{equation} 
and the asymmetry as
\begin{equation}
\delta_u=-\delta_{\eta}=0.839.
\end{equation}
We compare the resummed results for the partial width and branching fractions in Fig.~\ref{fig:Wpsi}, and the corresponding asymmetry in Fig.~\ref{fig:Asymmetry}, respectively. We see that overall shapes behave in a similar way, while corrections can be sizable.

\section{Nearly degenerate $\psi$ production}\label{app:BreitWigner}
When the mass difference of $\psi_1$ and $\psi_2$ is comparable with their widths, the Breit-Wigner effect needs to be considered, and the interference effect would enhance the cross-section of $\psi$ production at colliders.  In this appendix we calculate the tree-level interference effects of $\psi_1$ and $\psi_2$.

\subsection{Narrow width approximation}
We first briefly review the narrow width approximation (NWA) before considering near-degenerate $\psi_i$'s.  The amplitude of single $A\to\psi \to B$ production can be written as
\begin{equation}
    i\calM = y_B \calT_{B} \frac{\pslash - M_\psi}{p^2-M^2_\psi + i \Gamma M_\psi } y_{A} \calT_{A}
    = (y_A y_B) \calT_B (\pslash - M_\psi) \calT_A \frac{1}{D_{\psi}},
\end{equation}
where $\calT_B$ and $\calT_A$ are the vertices of $\psi$ production and decay, $y_B$ and $y_A$ are the corresponding coupling coefficients, and we have defined the propagator $D_\psi^{-1}=(p^2-M_\psi^2+i\Gamma M_\psi)^{-1}$.

When the width of $\psi$ is very narrow, \ie, $\Gamma\ll M_\psi$, the NWA gives a very accurate approach to subtract the on-shell $\psi$ contribution from the squared matrix element.  In the NWA, the denominator $|D_\psi|^{-2}$ in the squared amplitude is approximated as a delta function
\begin{equation}\label{eq:NWA}
 \begin{aligned} 
\int_0^\infty dp^2 \frac{1}{|D_\psi|^2}= \frac{\pi}{M_\psi\Gamma} 
\implies \frac{1}{|D_\psi|^2}\approx \frac{\pi}{M_\psi\Gamma} \delta(p^2-M_\psi^2).
\end{aligned}   
\end{equation}
Then the squared matrix element is
\begin{equation}
\begin{aligned}
    |\calM|^2 &= \Big( \calT_B (\pslash - M_\psi) \calT_A \Big)^\dagger \Big( \calT_B (\pslash - M_\psi) \calT_A \Big) \frac{|y_Ay_B|^2}{|D_\psi|^2} \\
    & = |y_Ay_B|^2\frac{\pi}{M_\psi \Gamma} \delta(p^2-M_\psi^2) \sum_{s,s'} \tr\left[ \calT_B u_s(p)\bar u_s(p) \calT_A \calT_A^\dagger u_{s'}(p)\bar u_{s'}(p) \calT_B^\dagger \right],
\end{aligned}    
\end{equation}
where $u_s(p)$ is the on-shell 4-component spinor of $\psi$ with spin $s$.

\subsection{Single $\psi$ production}
Same as before, we take $|y_{L1}|=|y_{L2}|$ and $|y_{R1}|=|y_{R2}|$ so that the widths of $\psi_1$ and $\psi_2$ are the same at tree level. We define a general mass splitting and average
\begin{equation}
\Delta M_{\psi}\equiv M_{\psi_1}-M_{\psi_2}=\alpha\Gamma,~
 M_{\psi}\equiv (M_{\psi_1}+M_{\psi_2})/2,
\end{equation}
where $\alpha=1/2$ corresponds to the maximal asymmetry condition in Eq.~(\ref{eq:dMpsi}).

At tree-level, the couplings of $\psi_1$ and $\psi_2$ share exactly the same Lorentz structures, and the only difference is the coupling strength. Therefore, with two generations of $\psi$'s, the single $\psi$ production amplitude is rewritten as
\begin{equation}
    i\calM_{1+2} = i\calM_1 + i\calM_2
    = (y_{A1} y_{B1}) \calT_B \frac{\pslash - M_{\psi_1}}{D_{1}} \calT_A + (y_{A2} y_{B2}) \calT_B \frac{\pslash - M_{\psi_2}}{D_{2}} \calT_A,    
\end{equation}
where $D_i^{-1}=(p^2-M_{\psi_i}^2+i\Gamma M_{\psi_i})^{-1}$.
Similar to Eq.~\eqref{eq:NWA}, the phase space integral of $|D_i|^{-2}$ contributes dominantly at $p^2\sim M_\psi^2$, and the propagator can be viewed as a delta function.  The key point is that the interference between the two propagators also has a considerable contribution when $\alpha\lesssim 1$:
\begin{equation}\label{eq:NWAD1D2}
\begin{aligned}
    \frac{1}{|D_i|^2} &\approx  \frac{\pi}{M_\psi\Gamma}\delta(p^2-M_\psi^2), \\
    \Re\left(\frac{1}{D_1^* D_2}\right) &\approx  \frac{\pi}{M_\psi\Gamma}\frac{1}{1+\alpha^2}\delta(p^2-M_\psi^2), \\
  \Im\left(\frac{1}{D_1^* D_2}\right) &\approx \frac{\pi}{M_\psi\Gamma}\frac{\alpha}{1+\alpha^2}\delta(p^2-M_\psi^2).
\end{aligned}    
\end{equation}
Therefore, the interference effect between $\psi_1$ and $\psi_2$ production is non-negligible. The squared matrix element of the process $A\to \psi_i \to B$ is
\begin{equation}
\begin{aligned}
    |\calM_{1+2}|^2 
    \approx \frac{\pi}{M_\psi \Gamma} \delta(p^2-M_\psi^2) \Bigg(|y_{A1}y_{B1}|^2  &\sum_{s,s'} \tr\left[ \calT_B u_s(p)\bar u_s(p) \calT_A \calT_A^\dagger u_{s'}(p)\bar u_{s'}(p) \calT_B^\dagger \right] \\
    +|y_{A2}y_{B2}|^2 &  \sum_{s,s'} \tr\left[ \calT_B u_s(p)\bar u_s(p) \calT_A \calT_A^\dagger u_{s'}(p)\bar u_{s'}(p) \calT_B^\dagger \right]  \\
    + \frac{2\Re(y_{A1}y_{B1}y_{A2}^*y_{B2}^*)}{1+\alpha^2}& \sum_{s,s'} \tr\left[ \calT_B u_s(p)\bar u_s(p) \calT_A \calT_A^\dagger u_{s'}(p)\bar u_{s'}(p) \calT_B^\dagger \right] \\
    + \frac{2\alpha\Im(y_{A1}y_{B1}y_{A2}^*y_{B2}^*)}{1+\alpha^2} &  \sum_{s,s'} \tr\left[ \calT_B u_s(p)\bar u_s(p) \calT_A \calT_A^\dagger u_{s'}(p)\bar u_{s'}(p) \calT_B^\dagger \right] \Bigg).
\end{aligned}
\end{equation}
In comparison with only one $\psi$, the squared amplitude of $A\to \psi_i \to B$ can be obtained with a scale factor.  Assuming that $|y_{A1}|=|y_{A2}|=y_A$ and $|y_{B1}|=|y_{B2}|=y_B$, we have
\begin{equation}
    |\calM_{1+2}|^2
    = \Bigg( 2 +\frac{2}{1+\alpha^2}\frac{\Re(y_{A1}y_{B1}y_{A2}^*y_{B2}^*)}{{|y_{A}y_{B}|^2}} + \frac{2\alpha}{1+\alpha^2} \frac{\Im(y_{A1}y_{B1}y_{A2}^*y_{B2}^*)}{{|y_{A}y_{B}|^2}} \Bigg)|\calM_1|^2.
\end{equation}

A single $\psi$ can be produced at hadron colliders through $ug \to \psi \phi$ and $\psi$ can decay to $u\phi$ or $N\eta$. 
For the single $\psi_i$ production with a mono-jet signal $ug\to \psi (\to u\phi) \phi$, using the benchmark parameter choice in Eq.~\eqref{eq:Y}, we have
\begin{equation}
\begin{aligned}
&\sum_i \sigma(ug\to \psi_i \phi \to u\phi\phi)
=2\left(1+\frac{1}{1+\alpha^2}\right) \sigma(ug\to \psi \phi \to u\phi\phi),\\
&\sum_i \sigma(ug\to \psi_i \phi \to N\eta\phi) 
=2\left(1+\frac{\alpha}{1+\alpha^2}\right) \sigma(ug\to \psi \phi \to N\eta\phi).    
\end{aligned}
\end{equation}

\begin{figure}
\includegraphics[width=0.49\textwidth]{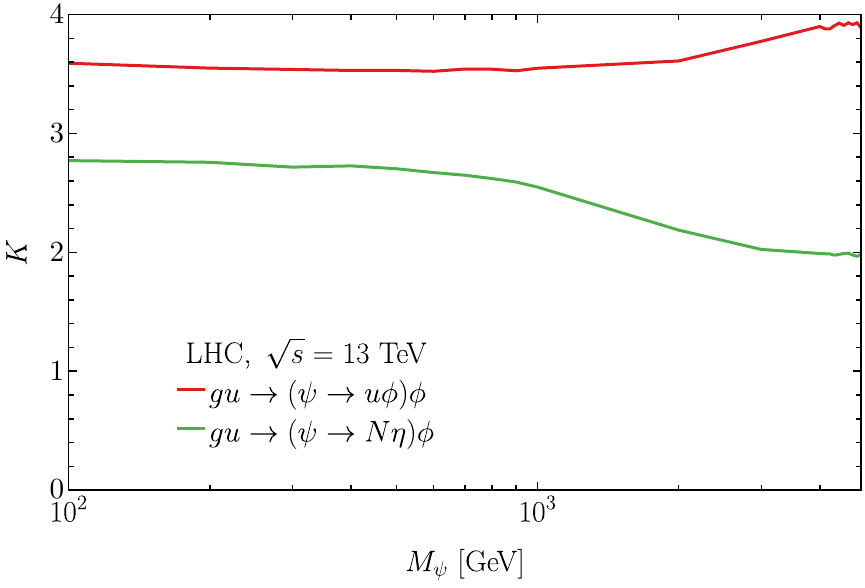}
\includegraphics[width=0.49\textwidth]{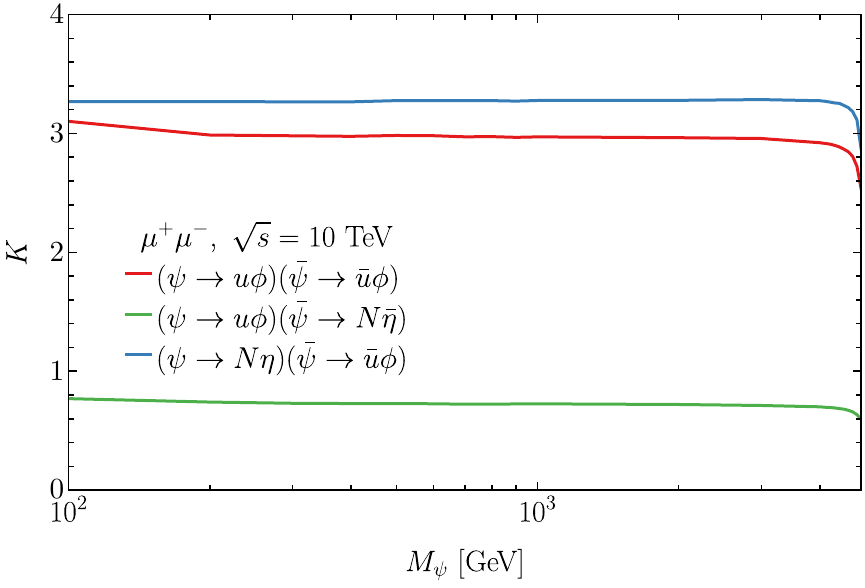}
\caption{The interference effect between $\psi_{1}$ and $\psi_{2}$ for the single and pair production at the 13 TeV LHC (left) and muon collider 10 TeV (right), respectively.}
\label{fig:intf}
\end{figure}

At tree-level, $\br^{(0)}(\psi\to ug)=\br^{(0)}(\psi\to N\eta)=0.5$ when $M_\eta, M_N, M_\phi \ll M_\psi$, and therefore $\sigma(ug\to \psi \phi \to u\phi\phi)=\sigma(ug\to \psi \phi \to N\eta\phi)$.  However, a ``decay asymmetry" of $\psi\to u\phi$ and $\psi \to N\eta$ arises due to the interference effect between $\psi_1$ and $\psi_2$ at tree level, which gives an enhancement $K$-factor of 3.6 and 2.8. In Fig.~\ref{fig:intf} (left), we present a numerical demonstration of the $K$-factors with the single $\psi$ production at the 13 TeV LHC. Besides the confirmation of our general estimation, a small variation occurs for $M_\psi\gtrsim1.5~\TeV$, as a consequence of the mass correction in the phase space.

\subsection{$\psi\bar\psi$ pair production}
In the case of $\psi\bar\psi$ pair production, the interference between $\psi_1$ and $\psi_2$ and the interference between $\bar\psi_1$ and $\bar \psi_2$ happen together. For simplicity, we use $D_i$($\bar D_i$) to denote the propagator of $\psi_i$($\bar \psi_i$), \ie,
\begin{equation}
  D_i^{-1} = \frac{1}{p^{2}-M_{\psi_i}^{2}+i \Gamma M_{\psi_i}},\qquad
  \bar D_i^{-1} = \frac{1}{\bar p^{2}-M_{\psi_i}^{2}+i \Gamma M_{\psi_i}}.
\end{equation}
Here $p$ and $\bar p$ are the four momentum of $\psi$ and $\bar \psi$ respectively.
The amplitude of the processes $I\to \psi_i\bar\psi_i \to  A\bar  B$ can be written as
\begin{equation}
    \calM_{1+2} \propto  \frac{y_{A1} y_{B1}^*}{D_1 \bar D_1} +  \frac{y_{A2} y_{B2}^*}{D_2 \bar D_2},
\end{equation}
where $A/\bar B$ are the decay products of $\psi/\bar \psi$, and the initial states $I$ can be  $e^+e^-$, $gg$ or $q\bar q$.
Similar to Eq.~(\ref{eq:NWAD1D2}), now we also need to consider the interference between $\bar D_1$ and $\bar D_2$, then the amplitude square is written as
\begin{equation}
\begin{aligned}
  |\mathcal{M}_{1+2}|^2&\propto \Bigg[ \frac{|y_{1A}y_{1B}^*|^2}{|D_1|^2 |\bar D_1 |^2}  + \frac{|y_{2A}y_{2B}^*|^2}{|D_2|^2 |\bar D_2 |^2} + 2 \Re\left( \frac{y_{A1}^* y_{B1} y_{A2} y_{B2}^*}{D_1^* \bar D_1^* D_2 \bar D_2} \right)  \Bigg] \\
  &=\Bigg[ \frac{|y_{1A}y_{1B}^*|^2}{|D_1|^2 |\bar D_1 |^2}  + \frac{|y_{2A}y_{2B}^*|^2}{|D_2|^2 |\bar D_2 |^2} \\   
  & \quad + 2 \Re(y_{1A}^* y_{1B} y_{2A} y_{2B}^*) \left( \Re(\frac{1}{D_1^* D_2}) \Re(\frac{1}{\bar D_1^* \bar D_2 }) - \Im(\frac{1}{D_1^* D_2}) \Im(\frac{1}{\bar D_1^* \bar D_2 }) \right) \\
  &\quad  - 2 \Im(y_{1A}^* y_{1B} y_{2A} y_{2B}^*) \left( \Re(\frac{1}{D_1^* D_2}) \Im(\frac{1}{\bar D_1^* \bar D_2 }) + \Im(\frac{1}{D_1^* D_2}) \Re(\frac{1}{\bar D_1^* \bar D_2}) \right) \Bigg].
\end{aligned}
\end{equation}
Substituting Eq.~(\ref{eq:NWAD1D2}) in the above equation, it is found that the amplitude squared is still proportional to $\delta(p^2-M_\psi^2)\delta(\bar p^2-M_\psi^2)$.  Then the ratio between $|\calM_{1+2}|^2$ and $|\calM_{1}|^2$ are given by
\begin{equation}
    |\calM_{1+2}|^2 
    = \Bigg(2+ \frac{2\Re(y_{A1}^* y_{B1} y_{A2} y_{B2}^*)}{|y_{A}|^2|y_{B}|^2} \frac{1-\alpha^2}{(1+\alpha^2)^2} - \frac{2\Im(y_{A1}^* y_{B1} y_{A2} y_{B2}^*)}{|y_{A}|^2|y_{B}|^2}\frac{2\alpha}{(1+\alpha^2)^2} \Bigg) |\calM_1|^2,
\end{equation}
where it is still assumed that $|y_{A1}|=|y_{A2}|=y_A$ and $|y_{B1}|=|y_{B2}|=y_B$.

When $\psi$ and $\bar \psi$ decay to the same final states, \ie, $A=B$, the scale factor is $2+2(1-\alpha^2)/(1+\alpha^2)^2\sim2.96$ with $\alpha=1/2$. If we consider the asymmetry between the $u+\slashed{E}_T$ and $\bar{u}+\slashed{E}_T$ in the $\psi\bar\psi$ pair production, and take the parameter assumption in Eq.~\eqref{eq:Y}, the corresponding cross-sections become
\begin{equation}
\begin{aligned}
\sum_{i=1,2}\sigma(\psi_i (\to u \phi) \bar \psi_i (\to N\bar\eta)) &= 2\left(1-\frac{2\alpha}{(1+\alpha^2)^2}\right)\sigma(\psi(\to u \phi) \bar \psi (\to N\bar \eta)), \\
\sum_{i=1,2}\sigma(\psi_i (\to N \eta) \bar \psi_i (\to \bar u \phi)) &= 2\left(1+\frac{2\alpha}{(1+\alpha^2)^2}\right)\sigma( \psi(\to N \eta) \bar \psi (\to \bar u \bar\phi)).
\end{aligned}
\end{equation}
As a consequence, the asymmetry from the interference is $2\alpha/(1+\alpha^2)^2=0.64$, with overall $K$-factors of 0.72 and 3.28. In Fig.~\ref{fig:intf} (right), we take the $\psi\bar{\psi}$ pair production at a 10 TeV muon collider for a numerical demonstration. It confirms our estimation, while a small variation occurs only around the threshold region where $M_\psi\sim\sqrt{s}/2$.

\bibliographystyle{utphys}
\bibliography{ref}

\providecommand{\href}[2]{#2}\begingroup\raggedright\begin{thebibliography}{10}

\bibitem{ParticleDataGroup:2020ssz}
{\bfseries Particle Data Group} Collaboration, P.~A. Zyla {\em et~al.},
  ``{Review of Particle Physics},''
  \href{http://dx.doi.org/10.1093/ptep/ptaa104}{{\em PTEP} {\bfseries 2020}
  no.~8, (2020) 083C01}.

\bibitem{Planck:2018vyg}
{\bfseries Planck} Collaboration, N.~Aghanim {\em et~al.}, ``{Planck 2018
  results. VI. Cosmological parameters},''
  \href{http://dx.doi.org/10.1051/0004-6361/201833910}{{\em Astron. Astrophys.}
  {\bfseries 641} (2020) A6}, \href{http://arxiv.org/abs/1807.06209}{{\ttfamily
  arXiv:1807.06209 [astro-ph.CO]}}. [Erratum: Astron.Astrophys. 652, C4
  (2021)].

\bibitem{Sakharov:1967dj}
A.~D. Sakharov, ``{Violation of CP Invariance, C asymmetry, and baryon
  asymmetry of the universe},''
  \href{http://dx.doi.org/10.1070/PU1991v034n05ABEH002497}{{\em Pisma Zh. Eksp.
  Teor. Fiz.} {\bfseries 5} (1967) 32--35}.

\bibitem{Ellis:1978xg}
J.~R. Ellis, M.~K. Gaillard, and D.~V. Nanopoulos, ``{Baryon Number Generation
  in Grand Unified Theories},''
  \href{http://dx.doi.org/10.1016/0370-2693(79)91190-0}{{\em Phys. Lett. B}
  {\bfseries 80} (1979) 360}. [Erratum: Phys.Lett.B 82, 464 (1979)].

\bibitem{Yoshimura:1978ex}
M.~Yoshimura, ``{Unified Gauge Theories and the Baryon Number of the
  Universe},'' \href{http://dx.doi.org/10.1103/PhysRevLett.41.281}{{\em Phys.
  Rev. Lett.} {\bfseries 41} (1978) 281--284}. [Erratum: Phys.Rev.Lett. 42, 746
  (1979)].

\bibitem{Weinberg:1979bt}
S.~Weinberg, ``{Cosmological Production of Baryons},''
  \href{http://dx.doi.org/10.1103/PhysRevLett.42.850}{{\em Phys. Rev. Lett.}
  {\bfseries 42} (1979) 850--853}.

\bibitem{Kolb:1979qa}
E.~W. Kolb and S.~Wolfram, ``{Baryon Number Generation in the Early
  Universe},'' \href{http://dx.doi.org/10.1016/0550-3213(82)90012-8}{{\em Nucl.
  Phys. B} {\bfseries 172} (1980) 224}. [Erratum: Nucl.Phys.B 195, 542 (1982)].

\bibitem{Fry:1980bd}
J.~N. Fry, K.~A. Olive, and M.~S. Turner, ``{Hierarchy of Cosmological Baryon
  Generation},'' \href{http://dx.doi.org/10.1103/PhysRevLett.45.2074}{{\em
  Phys. Rev. Lett.} {\bfseries 45} (1980) 2074}.

\bibitem{Fukugita:1986hr}
M.~Fukugita and T.~Yanagida, ``{Baryogenesis Without Grand Unification},''
  \href{http://dx.doi.org/10.1016/0370-2693(86)91126-3}{{\em Phys. Lett. B}
  {\bfseries 174} (1986) 45--47}.

\bibitem{Babu:2006xc}
K.~S. Babu, R.~N. Mohapatra, and S.~Nasri, ``{Post-Sphaleron Baryogenesis},''
  \href{http://dx.doi.org/10.1103/PhysRevLett.97.131301}{{\em Phys. Rev. Lett.}
  {\bfseries 97} (2006) 131301},
  \href{http://arxiv.org/abs/hep-ph/0606144}{{\ttfamily arXiv:hep-ph/0606144}}.

\bibitem{Babu:2006wz}
K.~S. Babu, R.~N. Mohapatra, and S.~Nasri, ``{Unified TeV Scale Picture of
  Baryogenesis and Dark Matter},''
  \href{http://dx.doi.org/10.1103/PhysRevLett.98.161301}{{\em Phys. Rev. Lett.}
  {\bfseries 98} (2007) 161301},
  \href{http://arxiv.org/abs/hep-ph/0612357}{{\ttfamily arXiv:hep-ph/0612357}}.

\bibitem{Allahverdi:2010im}
R.~Allahverdi, B.~Dutta, and K.~Sinha, ``{Baryogenesis and Late-Decaying
  Moduli},'' \href{http://dx.doi.org/10.1103/PhysRevD.82.035004}{{\em Phys.
  Rev. D} {\bfseries 82} (2010) 035004},
  \href{http://arxiv.org/abs/1005.2804}{{\ttfamily arXiv:1005.2804 [hep-ph]}}.

\bibitem{Allahverdi:2010rh}
R.~Allahverdi, B.~Dutta, and K.~Sinha, ``{Cladogenesis: Baryon-Dark Matter
  Coincidence from Branchings in Moduli Decay},''
  \href{http://dx.doi.org/10.1103/PhysRevD.83.083502}{{\em Phys. Rev. D}
  {\bfseries 83} (2011) 083502},
  \href{http://arxiv.org/abs/1011.1286}{{\ttfamily arXiv:1011.1286 [hep-ph]}}.

\bibitem{Allahverdi:2013tca}
R.~Allahverdi, B.~Dutta, R.~N. Mohapatra, and K.~Sinha, ``{A Supersymmetric
  Model for Dark Matter and Baryogenesis Motivated by the Recent CDMS
  Result},'' \href{http://dx.doi.org/10.1103/PhysRevLett.111.051302}{{\em Phys.
  Rev. Lett.} {\bfseries 111} (2013) 051302},
  \href{http://arxiv.org/abs/1305.0287}{{\ttfamily arXiv:1305.0287 [hep-ph]}}.

\bibitem{Allahverdi:2013mza}
R.~Allahverdi and B.~Dutta, ``{Natural GeV Dark Matter and the Baryon-Dark
  Matter Coincidence Puzzle},''
  \href{http://dx.doi.org/10.1103/PhysRevD.88.023525}{{\em Phys. Rev. D}
  {\bfseries 88} no.~2, (2013) 023525},
  \href{http://arxiv.org/abs/1304.0711}{{\ttfamily arXiv:1304.0711 [hep-ph]}}.

\bibitem{Allahverdi:2017edd}
R.~Allahverdi, P.~S.~B. Dev, and B.~Dutta, ``{A simple testable model of baryon
  number violation: Baryogenesis, dark matter, neutron\textendash{}antineutron
  oscillation and collider signals},''
  \href{http://dx.doi.org/10.1016/j.physletb.2018.02.019}{{\em Phys. Lett. B}
  {\bfseries 779} (2018) 262--268},
  \href{http://arxiv.org/abs/1712.02713}{{\ttfamily arXiv:1712.02713
  [hep-ph]}}.

\bibitem{Arcadi:2015ffa}
G.~Arcadi, L.~Covi, and M.~Nardecchia, ``{Gravitino Dark Matter and low-scale
  Baryogenesis},'' \href{http://dx.doi.org/10.1103/PhysRevD.92.115006}{{\em
  Phys. Rev. D} {\bfseries 92} no.~11, (2015) 115006},
  \href{http://arxiv.org/abs/1507.05584}{{\ttfamily arXiv:1507.05584
  [hep-ph]}}.

\bibitem{Baldo-Ceolin:1994hzw}
M.~Baldo-Ceolin {\em et~al.}, ``{A New experimental limit on neutron -
  anti-neutron oscillations},''
  \href{http://dx.doi.org/10.1007/BF01580321}{{\em Z. Phys. C} {\bfseries 63}
  (1994) 409--416}.

\bibitem{Super-Kamiokande:2011idx}
{\bfseries Super-Kamiokande} Collaboration, K.~Abe {\em et~al.}, ``{The Search
  for $n-\bar{n}$ oscillation in Super-Kamiokande I},''
  \href{http://dx.doi.org/10.1103/PhysRevD.91.072006}{{\em Phys. Rev. D}
  {\bfseries 91} (2015) 072006},
  \href{http://arxiv.org/abs/1109.4227}{{\ttfamily arXiv:1109.4227 [hep-ex]}}.

\bibitem{SNO:2017pha}
{\bfseries SNO} Collaboration, B.~Aharmim {\em et~al.}, ``{Search for
  neutron-antineutron oscillations at the Sudbury Neutrino Observatory},''
  \href{http://dx.doi.org/10.1103/PhysRevD.96.092005}{{\em Phys. Rev. D}
  {\bfseries 96} no.~9, (2017) 092005},
  \href{http://arxiv.org/abs/1705.00696}{{\ttfamily arXiv:1705.00696
  [hep-ex]}}.

\bibitem{Fridell:2021gag}
K.~Fridell, J.~Harz, and C.~Hati, ``{Probing baryogenesis with
  neutron-antineutron oscillations},''
  \href{http://dx.doi.org/10.1007/JHEP11(2021)185}{{\em JHEP} {\bfseries 11}
  (2021) 185}, \href{http://arxiv.org/abs/2105.06487}{{\ttfamily
  arXiv:2105.06487 [hep-ph]}}.

\bibitem{Baldes:2011mh}
I.~Baldes, N.~F. Bell, and R.~R. Volkas, ``{Baryon Number Violating Scalar
  Diquarks at the LHC},''
  \href{http://dx.doi.org/10.1103/PhysRevD.84.115019}{{\em Phys. Rev. D}
  {\bfseries 84} (2011) 115019},
  \href{http://arxiv.org/abs/1110.4450}{{\ttfamily arXiv:1110.4450 [hep-ph]}}.

\bibitem{Kuzmin:1985mm}
V.~A. Kuzmin, V.~A. Rubakov, and M.~E. Shaposhnikov, ``{On the Anomalous
  Electroweak Baryon Number Nonconservation in the Early Universe},''
\href{http://dx.doi.org/10.1016/0370-2693(85)91028-7}{{\em Phys. Lett.}
  {\bfseries 155B} (1985) 36}.

\bibitem{Chun:2017spz}
E.~J. Chun {\em et~al.}, ``{Probing Leptogenesis},''
  \href{http://dx.doi.org/10.1142/S0217751X18420058}{{\em Int. J. Mod. Phys. A}
  {\bfseries 33} no.~05n06, (2018) 1842005},
  \href{http://arxiv.org/abs/1711.02865}{{\ttfamily arXiv:1711.02865
  [hep-ph]}}.

\bibitem{Dick:1999je}
K.~Dick, M.~Lindner, M.~Ratz, and D.~Wright, ``{Leptogenesis with Dirac
  neutrinos},'' \href{http://dx.doi.org/10.1103/PhysRevLett.84.4039}{{\em Phys.
  Rev. Lett.} {\bfseries 84} (2000) 4039--4042},
  \href{http://arxiv.org/abs/hep-ph/9907562}{{\ttfamily arXiv:hep-ph/9907562}}.

\bibitem{Murayama:2002je}
H.~Murayama and A.~Pierce, ``{Realistic Dirac leptogenesis},''
  \href{http://dx.doi.org/10.1103/PhysRevLett.89.271601}{{\em Phys. Rev. Lett.}
  {\bfseries 89} (2002) 271601},
  \href{http://arxiv.org/abs/hep-ph/0206177}{{\ttfamily arXiv:hep-ph/0206177}}.

\bibitem{Minkowski:1977sc}
P.~Minkowski, ``{$\mu \to e\gamma$ at a Rate of One Out of $10^{9}$ Muon
  Decays?},'' \href{http://dx.doi.org/10.1016/0370-2693(77)90435-X}{{\em Phys.
  Lett. B} {\bfseries 67} (1977) 421--428}.

\bibitem{Yanagida:1979as}
T.~Yanagida, ``{HORIZONTAL SYMMETRY AND MASSES OF NEUTRINOS},''
{\em Conf. Proc.} {\bfseries C7902131} (1979) 95--99.

\bibitem{Gell-Mann:1979vob}
M.~Gell-Mann, P.~Ramond, and R.~Slansky, ``{Complex Spinors and Unified
  Theories},'' {\em Conf. Proc. C} {\bfseries 790927} (1979) 315--321,
  \href{http://arxiv.org/abs/1306.4669}{{\ttfamily arXiv:1306.4669 [hep-th]}}.

\bibitem{Mohapatra:1980yp}
R.~N. Mohapatra and G.~Senjanovic, ``{Neutrino Masses and Mixings in Gauge
  Models with Spontaneous Parity Violation},''
  \href{http://dx.doi.org/10.1103/PhysRevD.23.165}{{\em Phys. Rev. D}
  {\bfseries 23} (1981) 165}.

\bibitem{Schechter:1980gr}
J.~Schechter and J.~W.~F. Valle, ``{Neutrino Masses in SU(2) x U(1)
  Theories},'' \href{http://dx.doi.org/10.1103/PhysRevD.22.2227}{{\em Phys.
  Rev. D} {\bfseries 22} (1980) 2227}.

\bibitem{GAMBIT:2017gge}
{\bfseries GAMBIT} Collaboration, P.~Athron {\em et~al.}, ``{Status of the
  scalar singlet dark matter model},''
  \href{http://dx.doi.org/10.1140/epjc/s10052-017-5113-1}{{\em Eur. Phys. J. C}
  {\bfseries 77} no.~8, (2017) 568},
  \href{http://arxiv.org/abs/1705.07931}{{\ttfamily arXiv:1705.07931
  [hep-ph]}}.

\bibitem{Pilaftsis:1997jf}
A.~Pilaftsis, ``{CP violation and baryogenesis due to heavy Majorana
  neutrinos},'' \href{http://dx.doi.org/10.1103/PhysRevD.56.5431}{{\em Phys.
  Rev. D} {\bfseries 56} (1997) 5431--5451},
  \href{http://arxiv.org/abs/hep-ph/9707235}{{\ttfamily arXiv:hep-ph/9707235}}.

\bibitem{Gondolo:1990dk}
P.~Gondolo and G.~Gelmini, ``{Cosmic abundances of stable particles: Improved
  analysis},''
\href{http://dx.doi.org/10.1016/0550-3213(91)90438-4}{{\em Nucl. Phys.}
  {\bfseries B360} (1991) 145--179}.

\bibitem{LZ:2022lsv}
{\bfseries LZ} Collaboration, J.~Aalbers {\em et~al.}, ``{First Dark Matter
  Search Results from the LUX-ZEPLIN (LZ) Experiment},''
  \href{http://dx.doi.org/10.1103/PhysRevLett.131.041002}{{\em Phys. Rev.
  Lett.} {\bfseries 131} no.~4, (2023) 041002},
  \href{http://arxiv.org/abs/2207.03764}{{\ttfamily arXiv:2207.03764
  [hep-ex]}}.

\bibitem{DiMauro:2023tho}
M.~Di~Mauro, C.~Arina, N.~Fornengo, J.~Heisig, and D.~Massaro, ``{Dark matter
  in the Higgs resonance region},''
  \href{http://dx.doi.org/10.1103/PhysRevD.108.095008}{{\em Phys. Rev. D}
  {\bfseries 108} no.~9, (2023) 095008},
  \href{http://arxiv.org/abs/2305.11937}{{\ttfamily arXiv:2305.11937
  [hep-ph]}}.

\bibitem{Alwall:2014hca}
J.~Alwall, R.~Frederix, S.~Frixione, V.~Hirschi, F.~Maltoni, O.~Mattelaer,
  H.~S. Shao, T.~Stelzer, P.~Torrielli, and M.~Zaro, ``{The automated
  computation of tree-level and next-to-leading order differential cross
  sections, and their matching to parton shower simulations},''
  \href{http://dx.doi.org/10.1007/JHEP07(2014)079}{{\em JHEP} {\bfseries 07}
  (2014) 079}, \href{http://arxiv.org/abs/1405.0301}{{\ttfamily arXiv:1405.0301
  [hep-ph]}}.

\bibitem{Frederix:2018nkq}
R.~Frederix, S.~Frixione, V.~Hirschi, D.~Pagani, H.~S. Shao, and M.~Zaro,
  ``{The automation of next-to-leading order electroweak calculations},''
  \href{http://dx.doi.org/10.1007/JHEP11(2021)085}{{\em JHEP} {\bfseries 07}
  (2018) 185}, \href{http://arxiv.org/abs/1804.10017}{{\ttfamily
  arXiv:1804.10017 [hep-ph]}}. [Erratum: JHEP 11, 085 (2021)].

\bibitem{Kilian:2007gr}
W.~Kilian, T.~Ohl, and J.~Reuter, ``{WHIZARD: Simulating Multi-Particle
  Processes at LHC and ILC},''
  \href{http://dx.doi.org/10.1140/epjc/s10052-011-1742-y}{{\em Eur. Phys. J. C}
  {\bfseries 71} (2011) 1742}, \href{http://arxiv.org/abs/0708.4233}{{\ttfamily
  arXiv:0708.4233 [hep-ph]}}.

\bibitem{Moretti:2001zz}
M.~Moretti, T.~Ohl, and J.~Reuter, ``{O'Mega: An Optimizing matrix element
  generator},'' \href{http://arxiv.org/abs/hep-ph/0102195}{{\ttfamily
  arXiv:hep-ph/0102195}}.

\bibitem{Christensen:2010wz}
N.~D. Christensen, C.~Duhr, B.~Fuks, J.~Reuter, and C.~Speckner, ``{Introducing
  an interface between WHIZARD and FeynRules},''
  \href{http://dx.doi.org/10.1140/epjc/s10052-012-1990-5}{{\em Eur. Phys. J. C}
  {\bfseries 72} (2012) 1990}, \href{http://arxiv.org/abs/1010.3251}{{\ttfamily
  arXiv:1010.3251 [hep-ph]}}.

\bibitem{Degrande:2011ua}
C.~Degrande, C.~Duhr, B.~Fuks, D.~Grellscheid, O.~Mattelaer, and T.~Reiter,
  ``{UFO - The Universal FeynRules Output},''
  \href{http://dx.doi.org/10.1016/j.cpc.2012.01.022}{{\em Comput. Phys.
  Commun.} {\bfseries 183} (2012) 1201--1214},
  \href{http://arxiv.org/abs/1108.2040}{{\ttfamily arXiv:1108.2040 [hep-ph]}}.

\bibitem{Alloul:2013bka}
A.~Alloul, N.~D. Christensen, C.~Degrande, C.~Duhr, and B.~Fuks, ``{FeynRules
  2.0 - A complete toolbox for tree-level phenomenology},''
  \href{http://dx.doi.org/10.1016/j.cpc.2014.04.012}{{\em Comput. Phys.
  Commun.} {\bfseries 185} (2014) 2250--2300},
  \href{http://arxiv.org/abs/1310.1921}{{\ttfamily arXiv:1310.1921 [hep-ph]}}.

\bibitem{Staub:2013tta}
F.~Staub, ``{SARAH 4 : A tool for (not only SUSY) model builders},''
  \href{http://dx.doi.org/10.1016/j.cpc.2014.02.018}{{\em Comput. Phys.
  Commun.} {\bfseries 185} (2014) 1773--1790},
  \href{http://arxiv.org/abs/1309.7223}{{\ttfamily arXiv:1309.7223 [hep-ph]}}.

\bibitem{Hou:2019efy}
T.-J. Hou {\em et~al.}, ``{New CTEQ global analysis of quantum chromodynamics
  with high-precision data from the LHC},''
  \href{http://dx.doi.org/10.1103/PhysRevD.103.014013}{{\em Phys. Rev. D}
  {\bfseries 103} no.~1, (2021) 014013},
  \href{http://arxiv.org/abs/1912.10053}{{\ttfamily arXiv:1912.10053
  [hep-ph]}}.

\bibitem{Han:2021kes}
T.~Han, Y.~Ma, and K.~Xie, ``{Quark and gluon contents of a lepton at high
  energies},'' \href{http://dx.doi.org/10.1007/JHEP02(2022)154}{{\em JHEP}
  {\bfseries 02} (2022) 154}, \href{http://arxiv.org/abs/2103.09844}{{\ttfamily
  arXiv:2103.09844 [hep-ph]}}.

\bibitem{CLICdp:2018cto}
{\bfseries CLICdp, CLIC} Collaboration, T.~K. Charles {\em et~al.}, ``{The
  Compact Linear Collider (CLIC) - 2018 Summary Report},''
  \href{http://arxiv.org/abs/1812.06018}{{\ttfamily arXiv:1812.06018
  [physics.acc-ph]}}.

\bibitem{Accettura:2023ked}
C.~Accettura {\em et~al.}, ``{Towards a muon collider},''
  \href{http://dx.doi.org/10.1140/epjc/s10052-023-11889-x}{{\em Eur. Phys. J.
  C} {\bfseries 83} no.~9, (2023) 864},
  \href{http://arxiv.org/abs/2303.08533}{{\ttfamily arXiv:2303.08533
  [physics.acc-ph]}}. [Erratum: Eur.Phys.J.C 84, 36 (2024)].

\bibitem{Bartosik:2020xwr}
N.~Bartosik {\em et~al.}, ``{Detector and Physics Performance at a Muon
  Collider},'' \href{http://dx.doi.org/10.1088/1748-0221/15/05/P05001}{{\em
  JINST} {\bfseries 15} no.~05, (2020) P05001},
  \href{http://arxiv.org/abs/2001.04431}{{\ttfamily arXiv:2001.04431
  [hep-ex]}}.

\bibitem{Fermi:1924tc}
E.~Fermi, ``{On the Theory of the impact between atoms and electrically charged
  particles},'' \href{http://dx.doi.org/10.1007/BF03184853}{{\em Z. Phys.}
  {\bfseries 29} (1924) 315--327}.

\bibitem{vonWeizsacker:1934nji}
C.~F. von Weizsacker, ``{Radiation emitted in collisions of very fast
  electrons},'' \href{http://dx.doi.org/10.1007/BF01333110}{{\em Z. Phys.}
  {\bfseries 88} (1934) 612--625}.

\bibitem{Williams:1935dka}
E.~J. Williams, ``{Correlation of certain collision problems with radiation
  theory},'' {\em Kong. Dan. Vid. Sel. Mat. Fys. Med.} {\bfseries 13N4} no.~4,
  (1935) 1--50.

\bibitem{Buckley:2010fj}
M.~R. Buckley, B.~Echenard, D.~Kahawala, and L.~Randall, ``{Stable Colored
  Particles R-SUSY Relics or Not?},''
  \href{http://dx.doi.org/10.1007/JHEP01(2011)013}{{\em JHEP} {\bfseries 01}
  (2011) 013}, \href{http://arxiv.org/abs/1008.2756}{{\ttfamily arXiv:1008.2756
  [hep-ph]}}.

\bibitem{ATLAS:2019gqq}
{\bfseries ATLAS} Collaboration, M.~Aaboud {\em et~al.}, ``{Search for heavy
  charged long-lived particles in the ATLAS detector in 36.1 fb$^{-1}$ of
  proton-proton collision data at $\sqrt{s} = 13$ TeV},''
  \href{http://dx.doi.org/10.1103/PhysRevD.99.092007}{{\em Phys. Rev. D}
  {\bfseries 99} no.~9, (2019) 092007},
  \href{http://arxiv.org/abs/1902.01636}{{\ttfamily arXiv:1902.01636
  [hep-ex]}}.

\bibitem{Ghosh:2017vhe}
A.~Ghosh, T.~Mondal, and B.~Mukhopadhyaya, ``{Heavy stable charged tracks as
  signatures of non-thermal dark matter at the LHC : a study in some
  non-supersymmetric scenarios},''
  \href{http://dx.doi.org/10.1007/JHEP12(2017)136}{{\em JHEP} {\bfseries 12}
  (2017) 136}, \href{http://arxiv.org/abs/1706.06815}{{\ttfamily
  arXiv:1706.06815 [hep-ph]}}.

\bibitem{Schramm2017}
S.~Schramm, {\em Mono-jet Prospects at an Upgraded LHC},
  \href{http://dx.doi.org/10.1007/978-3-319-44453-6_8}{pp.~267--281}.
\newblock Springer International Publishing, Cham, 2017.
\newblock \url{https://doi.org/10.1007/978-3-319-44453-6_8}.

\bibitem{ATLAS:2017bfj}
{\bfseries ATLAS} Collaboration, M.~Aaboud {\em et~al.}, ``{Search for dark
  matter and other new phenomena in events with an energetic jet and large
  missing transverse momentum using the ATLAS detector},''
  \href{http://dx.doi.org/10.1007/JHEP01(2018)126}{{\em JHEP} {\bfseries 01}
  (2018) 126}, \href{http://arxiv.org/abs/1711.03301}{{\ttfamily
  arXiv:1711.03301 [hep-ex]}}.

\bibitem{ATLAS:2021kxv}
{\bfseries ATLAS} Collaboration, G.~Aad {\em et~al.}, ``{Search for new
  phenomena in events with an energetic jet and missing transverse momentum in
  $pp$ collisions at $\sqrt {s}$ =13 TeV with the ATLAS detector},''
  \href{http://dx.doi.org/10.1103/PhysRevD.103.112006}{{\em Phys. Rev. D}
  {\bfseries 103} no.~11, (2021) 112006},
  \href{http://arxiv.org/abs/2102.10874}{{\ttfamily arXiv:2102.10874
  [hep-ex]}}.

\bibitem{Bierlich:2022pfr}
C.~Bierlich {\em et~al.}, ``{A comprehensive guide to the physics and usage of
  PYTHIA 8.3}'' \href{http://dx.doi.org/10.21468/SciPostPhysCodeb.8}{{\em
  SciPost Phys. Codeb.} {\bfseries 2022} (2022) 8},
  \href{http://arxiv.org/abs/2203.11601}{{\ttfamily arXiv:2203.11601
  [hep-ph]}}.

\bibitem{CMS:2021tkn}
{\bfseries CMS} Collaboration, A.~M. Sirunyan {\em et~al.}, ``{Search for
  long-lived particles decaying to jets with displaced vertices in
  proton-proton collisions at $\sqrt{s}=$ 13 TeV},''
  \href{http://dx.doi.org/10.1103/PhysRevD.104.052011}{{\em Phys. Rev. D}
  {\bfseries 104} no.~5, (2021) 052011},
  \href{http://arxiv.org/abs/2104.13474}{{\ttfamily arXiv:2104.13474
  [hep-ex]}}.

\bibitem{ZurbanoFernandez:2020cco}
I.~Zurbano~Fernandez {\em et~al.}, ``{High-Luminosity Large Hadron Collider
  (HL-LHC): Technical design report},''.

\bibitem{ATLAS:2020xyo}
{\bfseries ATLAS} Collaboration, G.~Aad {\em et~al.}, ``{Search for long-lived,
  massive particles in events with a displaced vertex and a muon with large
  impact parameter in $pp$ collisions at $\sqrt{s} = 13$ TeV with the ATLAS
  detector},'' \href{http://dx.doi.org/10.1103/PhysRevD.102.032006}{{\em Phys.
  Rev. D} {\bfseries 102} no.~3, (2020) 032006},
  \href{http://arxiv.org/abs/2003.11956}{{\ttfamily arXiv:2003.11956
  [hep-ex]}}.

\bibitem{Farrar:1978xj}
G.~R. Farrar and P.~Fayet, ``{Phenomenology of the Production, Decay, and
  Detection of New Hadronic States Associated with Supersymmetry},''
  \href{http://dx.doi.org/10.1016/0370-2693(78)90858-4}{{\em Phys. Lett. B}
  {\bfseries 76} (1978) 575--579}.

\bibitem{ATLAS:1997ad}
{\bfseries ATLAS} Collaboration, ``{ATLAS muon spectrometer: Technical design
  report},''.

\bibitem{Han:2020uak}
T.~Han, Z.~Liu, L.-T. Wang, and X.~Wang, ``{WIMPs at High Energy Muon
  Colliders},'' \href{http://dx.doi.org/10.1103/PhysRevD.103.075004}{{\em Phys.
  Rev. D} {\bfseries 103} no.~7, (2021) 075004},
  \href{http://arxiv.org/abs/2009.11287}{{\ttfamily arXiv:2009.11287
  [hep-ph]}}.

\bibitem{Field:1977fa}
R.~D. Field and R.~P. Feynman, ``{A Parametrization of the Properties of Quark
  Jets},'' \href{http://dx.doi.org/10.1016/0550-3213(78)90015-9}{{\em Nucl.
  Phys. B} {\bfseries 136} (1978) 1}.

\bibitem{Krohn:2012fg}
D.~Krohn, M.~D. Schwartz, T.~Lin, and W.~J. Waalewijn, ``{Jet Charge at the
  LHC},'' \href{http://dx.doi.org/10.1103/PhysRevLett.110.212001}{{\em Phys.
  Rev. Lett.} {\bfseries 110} no.~21, (2013) 212001},
  \href{http://arxiv.org/abs/1209.2421}{{\ttfamily arXiv:1209.2421 [hep-ph]}}.

\bibitem{Li:2021uww}
H.~T. Li, B.~Yan, and C.~P. Yuan, ``{Jet charge: A new tool to probe the
  anomalous Zbb\textasciimacron{} couplings at the EIC},''
  \href{http://dx.doi.org/10.1016/j.physletb.2022.137300}{{\em Phys. Lett. B}
  {\bfseries 833} (2022) 137300},
  \href{http://arxiv.org/abs/2112.07747}{{\ttfamily arXiv:2112.07747
  [hep-ph]}}.

\bibitem{Li:2023tcr}
H.~T. Li, B.~Yan, and C.~P. Yuan, ``{Discriminating between Higgs Production
  Mechanisms via Jet Charge at the LHC},''
  \href{http://dx.doi.org/10.1103/PhysRevLett.131.041802}{{\em Phys. Rev.
  Lett.} {\bfseries 131} no.~4, (2023) 041802},
  \href{http://arxiv.org/abs/2301.07914}{{\ttfamily arXiv:2301.07914
  [hep-ph]}}.

\bibitem{Kang:2023ptt}
Z.-B. Kang, A.~J. Larkoski, and J.~Yang, ``{Towards a Nonperturbative
  Formulation of the Jet Charge},''
  \href{http://dx.doi.org/10.1103/PhysRevLett.130.151901}{{\em Phys. Rev.
  Lett.} {\bfseries 130} no.~15, (2023) 151901},
  \href{http://arxiv.org/abs/2301.09649}{{\ttfamily arXiv:2301.09649
  [hep-ph]}}.

\bibitem{Davidson:2008bu}
S.~Davidson, E.~Nardi, and Y.~Nir, ``{Leptogenesis},''
  \href{http://dx.doi.org/10.1016/j.physrep.2008.06.002}{{\em Phys. Rept.}
  {\bfseries 466} (2008) 105--177},
  \href{http://arxiv.org/abs/0802.2962}{{\ttfamily arXiv:0802.2962 [hep-ph]}}.

\bibitem{Pilaftsis:1998pd}
A.~Pilaftsis, ``{Heavy Majorana neutrinos and baryogenesis},''
  \href{http://dx.doi.org/10.1142/S0217751X99000932}{{\em Int. J. Mod. Phys. A}
  {\bfseries 14} (1999) 1811--1858},
  \href{http://arxiv.org/abs/hep-ph/9812256}{{\ttfamily arXiv:hep-ph/9812256}}.

\bibitem{Pilaftsis:2003gt}
A.~Pilaftsis and T.~E.~J. Underwood, ``{Resonant leptogenesis},''
  \href{http://dx.doi.org/10.1016/j.nuclphysb.2004.05.029}{{\em Nucl. Phys. B}
  {\bfseries 692} (2004) 303--345},
  \href{http://arxiv.org/abs/hep-ph/0309342}{{\ttfamily arXiv:hep-ph/0309342}}.

\end{thebibliography}\endgroup

\end{document}